\documentclass[journal]{IEEEtran}%
\usepackage{subfigure,psfrag,graphicx,amssymb,makeidx,multicol,footmisc,subfigure}
\usepackage[tbtags]{amsmath}
\usepackage{textcomp}
\usepackage[usenames,dvipsnames]{pstricks}
\usepackage{epsfig}
\usepackage{amsfonts}
\usepackage{amssymb}
\usepackage[noadjust]{cite}
\usepackage{graphicx}%
\usepackage{color}
\providecommand{\U}[1]{\protect\rule{.1in}{.1in}}
\DeclareMathOperator*{\diag}{diag}

\DeclareMathOperator*{\tr}{tr}

\newtheorem{definition}{Definition}

\newtheorem{lemma}{Lemma}
\newtheorem{remark}{Remark}

\include{MyBib.bib}
\begin{document}

\title{Linear Precoding and Equalization for Network MIMO with
Partial Cooperation}
\author{Saeed Kaviani, \IEEEmembership{Student Member, IEEE}, Osvaldo Simeone, \IEEEmembership{Senior Member, IEEE}, Witold A. Krzymie\'{n}, \IEEEmembership{Senior Member, IEEE} and Shlomo Shamai (Shitz), \IEEEmembership{Fellow, IEEE}
\thanks{Manuscript received 24\textsuperscript{th} June 2011; revised 14\textsuperscript{th} November 2011; accepted 5\textsuperscript{th} January 2012. The material in this paper was presented in part at the IEEE Global Communication Conference (GLOBECOM'11), Houston, U.S.A., Dec. 2011.
\newline \indent S. Kaviani and W. A. Krzymie\'{n} are with the Department of Electrical and Computer Engineering, University of Alberta, Edmonton, AB, Canada (e-mails: skaviani@ualberta.ca,wak@ece.ualberta.ca). They are also with \mbox{TRLabs}, Edmonton, AB, Canada. The work of S. Kaviani and W. A. Krzymie\'{n} was supported by TRLabs, Rohit Sharma Professorship, and the Natural Sciences and Engineering Research Council (NSERC) of Canada. \newline \indent O. Simeone is with the CWCSPR, New Jersey Institute of Technology, Newark, NJ, USA (e-mail: osvaldo.simeone@njit.edu). The work of O. Simeone was supported by
the U.S. N.S.F. under grant CCF-0914899. \newline \indent S. Shamai (Shitz) is with the Department of Electrical Engineering, Technion--Israel Institute of Technology, Haifa, Israel (e-mail: sshlomo@ee.technion.ac.il). The work of S. Shamai was supported by the Israel Science Foundation (ISF).}}
\maketitle

\begin{abstract}
A cellular multiple-input multiple-output (MIMO) downlink system is studied in
which each base station (BS) transmits to some of the users, so that each user receives its intended signal from a subset of the
BSs. This scenario is referred to as network MIMO\ with partial cooperation, since only a subset of the BSs are able to coordinate their transmission towards any user.
The focus of this paper is on the optimization of linear beamforming strategies at the BSs and
at the users for network MIMO with partial cooperation. Individual
power constraints at the BSs are enforced, along with constraints on the number of streams per user. It is first shown that the system
is equivalent to a MIMO\ interference channel with generalized linear
constraints (MIMO-IFC-GC). The problems of maximizing the sum-rate (SR) and
minimizing the weighted sum mean square error (WSMSE) of the data estimates are non-convex, and suboptimal solutions with reasonable complexity need to be devised. Based on this,
suboptimal techniques that aim at maximizing the sum-rate for the MIMO-IFC-GC
are reviewed from recent literature and extended to the MIMO-IFC-GC where necessary.
Novel designs that aim at minimizing the WSMSE are then proposed. 
Extensive numerical simulations are provided to compare the performance of the
considered schemes for realistic cellular systems.

\end{abstract}

\section{Introduction}\label{Sec_Introduction}

Interference is known to be major obstacle for realizing the spectral
efficiency increase promised by multiple-antenna techniques in wireless
systems. Indeed, the multiple-input multiple-output (MIMO) capacity gains are
severely degraded and limited in cellular environments due to the deleterious
effect of interference \cite{Dai04},\cite{Blum03}. Therefore, network-level
interference management appears to be of fundamental importance to overcome
this limitation and harness the gains of MIMO\ technology. Confirming this
point, multi-cell cooperation, also known as network MIMO, has been shown to significantly
improve the system performance \cite{Gesbert10}.

Network MIMO\ involves cooperative transmission by multiple base
stations (BSs) to each user. Depending on the level of multi-cell cooperation,
network MIMO reduces to a number of scenarios, ranging from a MIMO broadcast
channel (BC) \cite{Weingarten06} in case of full cooperation among all BSs, to
a MIMO interference channel \cite{Cadambe08},\cite{Maddah-Ali08} in case no
cooperation at the BSs is allowed. In general, network MIMO\ allows
cooperation only among a cluster of BSs for transmission to a certain user
\cite{Huang09,Zhang09} (see also references in \cite{Gesbert10}).

In this paper, we consider a MIMO interference channel with partial
cooperation at the BSs. It is noted that all BSs cooperating for transmission
to a certain user have to be informed about the message (i.e., the bit string)
intended for the user. This can be realized using the
backhaul links among the BSs and the central switching unit. We focus on the sum-rate maximization (SRM) and on the
minimization of weighted sum-MSE (WSMSE) under per-BS power constraints and constraints on the number of streams per user. Moreover, although
non-linear processing techniques such as vector precoding
\cite{Peel05,Muller08} may generally be useful, we focus on more practical
linear processing techniques. Both the SRM and WSMSE minimization (WSMMSE) problems are non-convex \cite{Boyd04}, and thus suboptimal design strategies of reasonable complexity are called for.

\par The contributions of this paper are as follows:
\begin{itemize}
\item[(i)] It is first shown in Sec.~\ref{sec_EquivalenceWithMIMOIFCGC} that network MIMO with partial BS cooperation, that is, with partial message knowledge, is equivalent to a MIMO interference channel in which each transmitter knows the message of only one user under generalized
linear constraints, which we refer to as MIMO-IFC-GC;
 \item[(ii)] We review the available suboptimal techniques that have been proposed for the SRM problem \cite{NgHuang10,Luo10,Liu10} and extend them to the MIMO-IFC-GC scenario where necessary in Sec.~\ref{Sec:PreviousTechniques}. Since these techniques are generally unable to enforce constraints on the number of streams, we also review and generalize techniques that are based on the idea of interference alignment \cite{Cadambe08} and are able to impose such constraints;
\item[(iii)] Then, we propose two novel suboptimal solutions for the WSMMSE problem in Sec.~\ref{Sec_MSEMinimization} under arbitrary constraints on the number of streams. It is noted that the WSMMSE problem without such constraints would be trivial, as it would result in zero MMSE and no stream transmitted. The proposed solutions are based on a novel insight into the single-user MMSE problem with multiple linear constraints, which is discussed in Sec.~\ref{sec_SingleUser};

\item[(iv)] Finally, extensive numerical simulations are provided in Sec.~\ref{sec_NumericalResults} to compare performance of the proposed schemes in realistic cellular systems.
\end{itemize}
Linear MMSE precoding and equalization techniques proposed in this paper were discussed briefly in \cite{Kaviani11a}. The detailed analysis and discussion (including the proofs to the lemmas) are included in this paper. Additionally, we have also reviewed and extended available solutions to the SRM problem. Furthermore, we have included discussions of the complexity and overhead of the proposed techniques and previously available (and/or extended) solutions.
\par \textit{Notation:} We denote the positive definite matrices as $\mathbf A \succeq \mathbf 0$. $\left[\cdot\right]^+$ denotes $\max(\cdot,0)$. Capital bold letters represent matrices and small bold letters represent vectors. We denote the transpose operator with $(\cdot)^\mathsf T$ and conjugate transpose (Hermitian) with $(\cdot)^\mathsf H$. $\mathbf A^{-\frac{1}{2}}$ represents the inverse square of positive definite matrix $\mathbf A$.

\section{System Model and Preliminaries}\label{sec_SystemModel}
%
\begin{figure}[b]
\centering
\scalebox{.8}
{\ \begin{pspicture}(0,-3.4929688)(6.5028124,3.4929688)
\psline[linewidth=0.02cm](1.7609375,-3.3054688)(4.5809374,-3.3054688)
\psline[linewidth=0.02cm](1.8009375,-1.3054688)(4.6009374,-2.3054688)
\psline[linewidth=0.02cm](1.8009375,-1.3054688)(4.6009374,-1.8654687)
\psline[linewidth=0.02cm](1.7809376,-1.2854687)(4.5809374,-0.26546875)
\psdots[dotsize=0.2](1.7809376,2.6945312)
\psdots[dotsize=0.2](1.8009375,1.5145313)
\psdots[dotsize=0.2](1.8009375,-1.3054688)
\psdots[dotsize=0.06](4.6009374,-1.1054688)
\psdots[dotsize=0.06](4.6009374,-1.2654687)
\psdots[dotsize=0.06](4.6009374,-1.4254688)
\psdots[dotsize=0.2](1.7809376,-3.3054688)
\psdots[dotsize=0.2](1.7809376,3.3145313)
\psline[linewidth=0.02cm](1.8209375,-2.7454689)(4.5609374,-3.2854688)
\psline[linewidth=0.02cm](1.7809376,-3.3054688)(4.6009374,-2.3054688)
\psdots[dotsize=0.06](1.7809376,-1.8854687)
\psdots[dotsize=0.06](1.7809376,-2.0454688)
\psdots[dotsize=0.06](1.7809376,-2.2054687)
\psdots[dotsize=0.06](4.5609374,-2.6854687)
\psdots[dotsize=0.06](4.5609374,-2.8454688)
\psdots[dotsize=0.06](4.5609374,-3.0054688)
\psdots[dotsize=0.2](1.8009375,-2.7254686)
\psdots[dotsize=0.2,fillstyle=solid,dotstyle=o](4.6009374,-0.26546875)
\psdots[dotsize=0.2,fillstyle=solid,dotstyle=o](4.6009374,-1.8654687)
\psdots[dotsize=0.2,fillstyle=solid,dotstyle=o](4.5809374,2.6945312)
\psdots[dotsize=0.2,fillstyle=solid,dotstyle=o](4.5809374,-3.2654688)
\psdots[dotsize=0.2,fillstyle=solid,dotstyle=o](4.6009374,-2.3054688)
\psline[linewidth=0.02cm](1.8409375,-1.2654687)(1.8609375,-1.3454688)
\psline[linewidth=0.02cm](1.8009375,-1.3054688)(4.6209373,-0.7054688)
\psellipse[linewidth=0.02,dimen=outer](4.6009374,-1.2854687)(0.3,1.26)
\usefont{T1}{ptm}{m}{n}
\rput(5.262344,-1.2154688){$\mathcal K_m$}
\usefont{T1}{ptm}{m}{n}
\rput(1.3823438,-1.3154688){$m$}
\psdots[dotsize=0.2](1.8009375,-0.32546875)
\psdots[dotsize=0.2,fillstyle=solid,dotstyle=o](4.6009374,-0.7054688)
\psline[linewidth=0.02cm](1.8009375,1.5145313)(4.6009374,0.71453124)
\psline[linewidth=0.02cm](1.8009375,-0.32546875)(4.5809374,0.71453124)
\psline[linewidth=0.02cm](1.8409375,0.17453125)(4.6009374,0.71453124)
\psdots[dotsize=0.2](1.8009375,0.17453125)
\psdots[dotsize=0.06](1.8009375,0.9945313)
\psdots[dotsize=0.06](1.8009375,0.83453125)
\psdots[dotsize=0.06](1.8009375,0.6745312)
\psellipse[linewidth=0.02,dimen=outer](1.8209375,0.5545313)(0.3,1.26)
\usefont{T1}{ptm}{m}{n}
\rput(4.8,3.8){Users}
\usefont{T1}{ptm}{m}{n}
\rput(1.8,3.8){BSs}
\usefont{T1}{ptm}{m}{n}
\rput(4.8523436,0.7445313){$k$}
\usefont{T1}{ptm}{m}{n}
\rput(1.1423438,0.5445312){$\mathcal M_k$}
\psdots[dotsize=0.06](4.5609374,1.8745313)
\psdots[dotsize=0.06](4.5609374,1.7145313)
\psdots[dotsize=0.06](4.5609374,1.5545312)
\usefont{T1}{ptm}{m}{n}
\rput(4.912344,3.3045313){$1$}
\usefont{T1}{ptm}{m}{n}
\rput(4.912344,2.7045312){$2$}
\usefont{T1}{ptm}{m}{n}
\rput(4.952344,-3.2954688){$K$}
\usefont{T1}{ptm}{m}{n}
\rput(1.4523437,3.3045313){$1$}
\usefont{T1}{ptm}{m}{n}
\rput(1.4523437,2.6845312){$2$}
\usefont{T1}{ptm}{m}{n}
\rput(1.4223437,-3.3154688){$M$}
\psdots[dotsize=0.06](1.7609375,2.3145313)
\psdots[dotsize=0.06](1.7609375,2.1545312)
\psdots[dotsize=0.06](1.7609375,1.9945313)
\psline[linewidth=0.02cm](1.7609375,3.3145313)(4.5809374,3.2945313)
\psline[linewidth=0.02cm](1.7809376,2.7145312)(4.5809374,3.2745314)
\psline[linewidth=0.02cm](1.7809376,2.7145312)(2.4209375,1.7145313)
\psline[linewidth=0.02cm](1.8209375,3.2745314)(3.0009375,2.6545312)
\psline[linewidth=0.02cm](1.8009375,2.6945312)(2.5009375,2.2545311)
\psline[linewidth=0.02cm](1.7809376,3.2945313)(2.6809375,2.4745312)
\psdots[dotsize=0.06](3.1609375,2.2945313)
\psdots[dotsize=0.06](3.1609375,1.5145313)
\psdots[dotsize=0.06](3.1609375,1.9145312)
\psdots[dotsize=0.2,fillstyle=solid,dotstyle=o](4.5809374,3.2945313)
\psdots[dotsize=0.2,fillstyle=solid,dotstyle=o](4.6009374,0.71453124)
\end{pspicture}
}\caption{A downlink model with partial BS cooperation or equivalently partial message knowledge.}%
\end{figure}
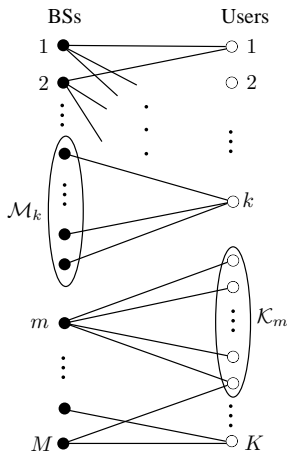

Consider the MIMO downlink system illustrated in Fig. 1
with $M$ base stations
(BSs) forming a set $\mathcal{M}$, and $K$ users forming a set $\mathcal{K}$.
Each BS is equipped with $n_{t}$ transmit antennas and each mobile user
employs $n_{r}$ receive antennas. The $m$th BS is provided with the messages
of its assigned users set $\mathcal{K}_{m}\subseteq\mathcal{K}$. In other
words, the $k$th user receives its message from a subset of $M_{k} $ BSs
$\mathcal{M}_{k}\subseteq\mathcal{M}$. Notice that, if $\mathcal{K}_{m}$
contains one user for each transmitter $m$ and $M_{k}=1,$ then the model at
hand reduces to a standard MIMO interference channel. Moreover, when all
transmitters cooperate in transmitting to all the users, i.e., $\mathcal{K}%
_{m}=\mathcal{K}$ for all $m\in\mathcal{M}$ or equivalently $M_{k}=M$, then we
have a MIMO broadcast channel (BC).

We now detail the signal model for the channel at hand, which is referred to
as \textit{MIMO\ interference channel with partial message sharing}. Define as
$\mathbf{u}_{k}=[u_{k,1}\ldots u_{k,d_{k}}]^{\mathsf{T}}\in\mathbb{C}^{d_{k}}$
the $d_{k}\times1$ complex vector representing the $d_{k}\leq\min(M_{k}%
n_{t},n_{r})$ independent information streams intended for user $k$. 
We assume that $\mathbf{u}_{k}\sim\mathcal{CN}(0,\mathbf{I}%
)$. The data streams $\mathbf u_k$ are known to all the BS in the set $\mathcal{M}_{k}.$ In
particular, if $m\in\mathcal{M}_{k},$ the $m$th BS precodes vector
$\mathbf{u}_{k}$ via a matrix $\mathbf{B}_{k,m}\in\mathbb{C}^{n_{t}\times
d_{k}}$, so that the signal $\tilde{\mathbf{x}}_{m}\in\mathbb{C}^{n_{t}}$ sent
by the $m$th BS can be expressed as%
\begin{equation}
\tilde{\mathbf{x}}_{m}=\sum\limits_{k\in\mathcal{K}_{m}}\mathbf{B}_{k,m}%
\mathbf{u}_{k}.
\end{equation}
Imposing a per-BS power constraint, the following constraint must be then
satisfied
\begin{align}\label{eqn:primary_powerConstraints}
\mathbb E\left[||\tilde{\mathbf{x}}_{m}||^{2}\right]  &  =\tr\left\{\mathbb E\left[\tilde{\mathbf{x}}%
_{m}\tilde{\mathbf{x}}_{m}^{\mathsf{H}}\right]\right\}\\
&  =\sum\limits_{k\in\mathcal{K}_{m}}\tr\left\{  \mathbf{B}_{k,m}\mathbf{B}%
_{k,m}^{\mathsf{H}}\right\}  \leq P_{m}, m=1,\ldots,M,\nonumber
\end{align}
where $P_{m}$ is the power constraint of the $m$th BS.

The signal received at the $k$th user can be written as
\begin{subequations}
\begin{align}
\mathbf{y}_{k}  &  =\sum\limits_{m=1}^{M}\widetilde{\mathbf{H}}_{k,m}\tilde
{\mathbf{x}}_{m}+\tilde{\mathbf{n}}_{k}\label{eqn:received_signal0}\\
&  =\sum\limits_{m\in\mathcal{M}_{k}}\widetilde{\mathbf{H}}_{k,m}\mathbf{B}%
_{k,m}\mathbf{u}_{k}+\sum\limits_{l\neq k}\sum\limits_{j\in\mathcal{M}_{l}}\widetilde
{\mathbf{H}}_{k,j}\mathbf{B}_{l,j}\mathbf{u}_{l}+\tilde{\mathbf{n}}_{k},
\label{eqn:received_signal01}%
\end{align}
where $\widetilde{\mathbf{H}}_{k,m}\in\mathbb{C}^{n_{r}\times n_{t}}$ is the
channel matrix between the $m$th BS and $k$th user and $\mathbf{n}_{k}$ is
additive complex Gaussian noise $\tilde{\mathbf{n}}_{k}\sim\mathcal{CN}%
(\mathbf{0},\mathbf{I})$\footnote{In case the noise is not uncorrelated across
the antennas, each user can always whiten it as a linear pre-processing step.
Therefore, a spatially uncorrelated noise can be assumed without loss of
generality.}$\mathbf{.}$ We assume ideal channel state information at all nodes. In (\ref{eqn:received_signal01}), we have
distinguished between the first term, which represents useful signal, the
second term, which accounts for interference, and the noise.

\subsection{Equivalence with MIMO-IFC-GC}\label{sec_EquivalenceWithMIMOIFCGC}

We now show that the MIMO\ interference channel with \textit{partial message
sharing} \textit{and} \textit{per-BS power constraints} described above is
equivalent to a specific MIMO\ interference channel with \textit{individual
message knowledge and generalized linear constraints,} which we refer to as
MIMO-IFC-GC.
\end{subequations}
\begin{definition}
(\textit{MIMO-IFC-GC}) The MIMO-IFC-GC consists of $K$ transmitters and $K$
receiver, where the $k$th transmitter has $m_{t,k}$ antennas and the $k$th
receiver has $m_{r,k}$ antennas. The received signal at the $k$th receiver is
\begin{equation}
\mathbf{y}_{k}=\mathbf{H}_{k,k}\mathbf{x}_{k}+\sum\limits_{l\neq k}\mathbf{H}%
_{k,l}\mathbf{x}_{l}+\mathbf{n}_{k},
\end{equation}
where $\mathbf{n}_{k}\sim\mathcal{CN}(\mathbf{0},\mathbf{I}),$ the inputs are
$\mathbf{x}_{k}\in \mathbb{C}^{m_{t,k}}$ and the channel matrix between the
$l$th transmitter and the $k$th receiver is $\mathbf{H}_{k,l}\in
\mathbb{C}^{m_{r,k}\times m_{t,k}}.$ The data vector intended for user $k$ is
$\mathbf{u}_{k}\in \mathbb{C}^{d_{k}}$ with $d_{k}\leq\min(m_{t,k},m_{r,k})$
and $\mathbf{u}_{k}\sim\mathcal{CN}(0,\mathbf{I}).$ The precoding matrix for
user $k$ is defined as $\mathbf{B}_{k}\in\mathbb{C}^{m_{t,k}\times d_{k}}$ so
that $\mathbf{x}_{k}=\mathbf{B}_{k}\mathbf{u}_{k}.$ The inputs $\mathbf{x}%
_{k}$ have to satisfy $M$ generalized linear constraints
\begin{equation}
\sum\limits_{k=1}^{K}\tr\left\{\mathbf{\Phi}_{k,m}\mathbb{E}\left[\mathbf{x}_{k}%
\mathbf{x}_{k}^{\mathsf{H}}\right]\right\}=\sum\limits_{k=1}^{K}\tr\left\{\mathbf{\Phi}_{k,m}\mathbf{B}_{k}\mathbf{B}%
_{k}^{\mathsf{H}}\right\}\leq P_{m},\label{eq:GeneralizedPowerConstraints}
\end{equation}
for given weight matrices $\mathbf{\Phi}_{k,m}\in \mathbb{C}^{m_{t,k}\times
m_{t,k}}$ and $m=1,\ldots,M.$ The weight matrices are such that matrices
$\sum_{m=1}^{M}\mathbf{\Phi}_{k,m}$ are positive definite for all
$k=1,\ldots,K.$
\label{def:IFC_GC}
\end{definition}

We remark that the positive definiteness of matrices $\sum
_{m=1}^{M}\mathbf{\Phi}_{k,m}$ guarantees that the system is not allowed
to transmit infinite power in any direction \cite{Liu10a}.

\begin{lemma}
\label{Lem:MIMO-IFC-GC}
Let $(l)_{k}$ be the $l$th BS in subset $\mathcal{M}%
_{k}$ of BSs that know user $k$'s message. The MIMO\ interference channel with partial message sharing (and per-transmitter power constraints) is equivalent to a
MIMO-IFC-GC. This equivalent MIMO-IFC-GC is defined with $m_{t,k}=M_{k}n_{t},$ $m_{r,k}=n_{r},$ channel matrices
\begin{equation}
\mathbf{H}_{k,l}=\left[  \widetilde{\mathbf{H}}_{k,(1)_{l}}\cdots
\widetilde{\mathbf{H}}_{k,(M_{l})_{l}}\right],
\end{equation}
beamforming matrices
\begin{equation}
\mathbf{B}_{k}=\left[  \mathbf{B}_{k,(1)_{k}}^{\mathsf{T}}\cdots
\mathbf{B}_{k,(M_{k})_{k}}^{\mathsf{T}}\right]  ^{\mathsf{T}}%
\label{eqn:BeamformingMatrix}%
\end{equation}
and weight matrices $\mathbf \Phi_{k,m}$ being all zero except that their $l$th $n_{t}\times n_{t}$ submatrix on the main diagonal is $\mathbf I_{n_t}$,  if $m=(l)_{k}$ (If $k\notin \mathcal{K}_{m}$ then $\mathbf{\Phi}_{k,m}=\mathbf{0}$). We emphasize that the definition of MIMO-IFC-GC and this equivalence rely on the assumption of linear processing at the transmitters.
\end{lemma}

\begin{IEEEproof}: The proof follows by inspection. Notice that matrices
$\sum_{m=1}^{M}\mathbf{\Phi}_{k,m}$ are positive definite by construction.
\end{IEEEproof}

Given the generality of the MIMO-IFC-GC, which includes the scenario of
interest of MIMO\ interference channel with partial message sharing as per the
Lemma above, in the following we focus on the MIMO-IFC-GC as defined above
and return to the cellular application in Sec. \ref{sec_NumericalResults}. It
is noted that a model that subsumes the MIMO-IFC-GC has been studied in \cite{Liu10a}, as discussed below.

\subsection{Linear Receivers and Mean Square Error}\label{sec_LinearReceiversAndMSE}

In this paper, we focus on the performance of the MIMO-IFC-GC under linear
processing at the receivers. Therefore, the $k$th receiver estimates the
intended vector $\mathbf{u}_{k}$ using the receive processing (or
equalization) matrix $\mathbf{A}_{k}\in\mathbb{C}^{d_{k}\times m_{r,k}}$ as
\begin{equation}
\mathbf{\hat{u}}_{k}=\mathbf{A}_{k}^{\mathsf{H}}\mathbf{y}_{k}.
\label{eqn:Estimate}%
\end{equation}

The most common performance measures, such as weighted sum-rate or bit error
rate, can be derived from the estimation error covariance matrix for each user
$k,$
\begin{equation}
\mathbf{E}_{k}=\mathbb{E}\left[  \left(  \mathbf{\hat{u}}_{k}%
-\mathbf{u}_{k}\right)  \left(  \mathbf{\hat{u}}_{k}-\mathbf{u}%
_{k}\right)  ^{\mathsf{H}}\right]  , \label{eqn:MSEdefined}%
\end{equation}
which is referred to as \textit{Mean Square Error (MSE)-matrix} (see \cite{Palomar06} for a review). The name comes from the fact that that the $j$th
term on the main diagonal of $\mathbf{E}_{k}$ is the MSE
\begin{equation}
\mathsf {MSE}_{k,j}=\mathbb{E[}\left\vert \hat{u}_{k,j}-u_{k,j}\right\vert ^{2}]
\label{eqn:MSE-user}%
\end{equation}
on the estimation of the $k$th user's $j$th data stream $u_{k,j}$. Using the
definition of MIMO-IFC-GC, it is easy to see that the MSE-matrix can be
written as a function of the equalization matrix $\mathbf{A}_{k}$ and \textit{all}
the transmit matrices $\{\mathbf{B}_{k}\}_{k=1}^{K}$ as
\begin{equation}%
\begin{split}
\mathbf{E}_{k}=  &  \mathbf{A}_{k}^{\mathsf{H}}\mathbf{H}_{k}\mathbf{B}%
_{k}\mathbf{B}_{k}^{\mathsf{H}}\mathbf{H}_{k,k}^{\mathsf{H}}\mathbf{A}%
_{k}-\mathbf{A}_{k}^{\mathsf{H}}\mathbf{H}_{k,k}\mathbf{B}_{k}\\
&-\mathbf{B}_{k}^{\mathsf{H}}\mathbf{H}_{k,k}^{\mathsf{H}}\mathbf{A}_{k}
+\mathbf{A}_{k}^{\mathsf{H}}\mathbf{\Omega}_{k}\mathbf{A}_{k}%
+\mathbf{I}_{k}.%
\end{split}
\label{eqn:MSE-matrix2}%
\end{equation}
where $\mathbf{\Omega}_{k}$ is the covariance matrix that accounts for noise
and interference at user $k$
\begin{equation}
\mathbf{\Omega}_{k}=\mathbf{I}+\sum\limits_{l\neq k}\mathbf{H}_{k,l}\mathbf{B}%
_{l}\mathbf{B}_{l}^{\mathsf{H}}\mathbf{H}_{k,l}^{\mathsf{H}}.
\label{eqn:Omega}%
\end{equation}

\section{Problem Definition and Preliminaries\label{Sec:ProblemDefinition}}

In this paper, we consider the optimization of the sum of some specific
functions $f_{k}\left(  \mathbf{E}_{k}\right)  $ of the MSE-matrices
$\mathbf{E}_{k}$ of all users $k=1,\ldots,K$ for the MIMO-IFC-GC. Specifically, we address the
following constrained optimization problem
\begin{equation}%
\begin{array}
[c]{rl}%
\underset{\mathbf{B}_{k},\mathbf{A}_{k},\forall k}%
{\text{minimize}} & \sum\limits_{k=1}^{K}f_{k}(\mathbf{E}_{k})\\
\text{subject to} & \sum\limits_{k=1}^{K}\tr\left\{  \mathbf{\Phi}_{k,m}%
\mathbf{B}_{k}\mathbf{B}_{k}^{\mathsf{H}}\right\}  \leq P_{m},
m=1,\ldots,M,
\end{array} \label{eqn:optimization11}%
\end{equation}
where the optimization is over all transmit beamforming matrices
$\mathbf{B}_{k}$ and equalization matrices $\mathbf{A}_{k}$.
Specifically, we focus on the \textit{weighted sum-MSE functions} (WSMSE)
\begin{equation}
f_{k}\left(  \mathbf{E}_{k}\right)  =\tr\left\{  \mathbf{W}_{k}\mathbf{E}%
_{k}\right\}  =\sum\limits_{j=1}^{d_{k}}w_{kj}\mathsf{MSE}_{kj}
\label{eqn:WeightedMSEfunction}%
\end{equation}
with given diagonal weight matrices $\mathbf{W}_{k}\in\mathbb{C}^{d_{k}\times
d_{k}}$ where the main diagonal of $\mathbf{W}_{k}$ is given by $[w_{k,1}%
,...,w_{k,d_{k}}]$ with non-negative weights $w_{kj}\geq0.$ With cost function
(\ref{eqn:WeightedMSEfunction}), we refer to the problem
(\ref{eqn:optimization11}) as the \textit{weighted sum-MSE} minimization
(\textit{WSMMSE}) problem.

Of more direct interest for communications systems is the maximization of the
sum-rate. This is obtained from (\ref{eqn:optimization11}) by selecting the
\textit{sum-rate (SR) functions}
\begin{equation}
f_{k}(\mathbf{E}_{k})=\log\left\vert \mathbf{E}_{k}\right\vert .
\label{eqn:RateFunction}%
\end{equation}
With cost function (\ref{eqn:RateFunction}), problem
(\ref{eqn:optimization11}) is referred to as the \textit{sum-rate
maximization} (SRM) problem. In fact, from information-theoretic
considerations, it can be seen that (\ref{eqn:RateFunction}) is the maximum
achievable rate (in bits per channel use) for the $k$th user where the signals
of the other users are treated as noise (see, e.g., \cite{Palomar06}).

\begin{remark}
Consider an iterative algorithm where at each iteration a WSMMSE problem is solved with the weight matrices $\mathbf W_k$ assumed to be non-diagonal and selected based on the previous MSE-matrix $\mathbf E_k$. This algorithm can approximate the solution of (\ref{eqn:optimization11}) for any general cost function $f_{k}%
(\mathbf{E}_{k})$. This was first pointed out in \cite{Christensen08} for the
weighted SRM problem in a MIMO BC, then in \cite{Schmidth09} for the single-antenna
interference channel and a general utility function, and has been generalized to a MIMO (broadcast) interference channel in \cite{Luo11} with conventional power constraints. It is not difficult to see that this result extends also to the MIMO-IFC-GC, which is not subsumed in the model of \cite{Luo11} due to the generalized linear constraints. We explicitly state this conclusion below.
\end{remark}
\par \textit{Lemma 2 \cite{Luo11}}: For strictly concave utility functions $f_k(\cdot)$ for all $k$, the global optimal solution of problem (\ref{eqn:optimization11}) and the solution of
\begin{equation}%
\begin{array}
[c]{rl}%
\underset{\mathbf{B}_{k},\mathbf{A}_{k},\mathbf W_k, \forall k}%
{\text{minimize}} & \sum\limits_{k=1}^{K}\left\{\tr\left\{\mathbf W_k\mathbf E_k\right\}-\tr\left\{\mathbf W_kg_k(\mathbf W_k)\right\}\right. \\ & \left.+ f_{k}(g_k(\mathbf{W}_{k}))\right\}\\
\text{subject to} & \sum\limits_{k=1}^{K}\tr\left\{  \mathbf{\Phi}_{k,m}%
\mathbf{B}_{k}\mathbf{B}_{k}^{\mathsf{H}}\right\}  \leq P_{m},
m=1,\ldots,M,
\end{array} \label{eqn:optimization11a}%
\end{equation}
where $g_k(\cdot)$ is the inverse function of the $\nabla f_{k}(\cdot)$, are the same.

Consequently, in order to find an approximate solution of (\ref{eqn:optimization11}), at each step matrices $\mathbf W_k$ for $k=1.\ldots,K$ are updated by solving (\ref{eqn:optimization11a}) with respect to $\mathbf W_k$ only (i.e., we keep $(\mathbf A_k,\mathbf B_k), \forall k$ unchanged in this step). Then, using the obtained matrices $\mathbf W_k$, for $k=1,\ldots,K$, the problem (\ref{eqn:optimization11a}) reduces to a WSMMSE problem with respect to matrices $\mathbf A_k$ and $\mathbf B_k$ for $k=1,\ldots,K$ (i.e., matrices $\mathbf W_k$ are kept fixed). This results in the iterative algorithm, that is discussed in Remark 1 and that leads to a suboptimal solution of (\ref{eqn:optimization11}). 
In the special case of the SRM problem, we have $f_k(\mathbf E_k)=\log\left|\mathbf E_k\right|$ and $g_k(\mathbf W_k)=\mathbf W_k^{-1}$, in which problem (\ref{eqn:optimization11a}) is then equivalent to the problem
\begin{equation}%
\begin{array}
[c]{rl}%
\underset{\mathbf{B}_{k},\mathbf{A}_{k},\mathbf W_k, \forall k}%
{\text{minimize}} & \sum\limits_{k=1}^{K}\tr\left\{\mathbf W_k\mathbf E_k\right\}-\sum\limits_{k=1}^K\log \left|\mathbf W_k\right|\\
\text{subject to} & \sum\limits_{k=1}^{K}\tr\left\{  \mathbf{\Phi}_{k,m}%
\mathbf{B}_{k}\mathbf{B}_{k}^{\mathsf{H}}\right\}  \leq P_{m},
m=1,\ldots,M.
\end{array}
 \label{eqn:optimization11b}%
\end{equation}
The optimization problem (\ref{eqn:optimization11b}) can be solved in an iterative fashion, where at each iteration the weights are selected as $\mathbf W_k^{\star}=\mathbf E_k^{-1}$ and then the WSMMSE problem is solved with respect to matrices $(\mathbf A_k,\mathbf B_k)$ for $k=1,\ldots,K$.

\section{The Single-User Case ($K=1$)}\label{sec_SingleUser}
The WSMMSE and SRM problems are non-convex and thus global optimization is
generally prohibitive. In this section, we address the case of a single user
($K=1).$ In particular, the SRM\ problem with $K=1$ is non-convex if one includes constraints on the number of streams $d_1$, but is otherwise convex and in this special case can be solved efficiently \cite{Palomar06}. The global optimal solution for the single-user problem with multiple linear power constraint (and a rank constraint) is still unknown \cite{YuLau11}. The WSMMSE problem is trivial without rank constraint, as explained above, and is non-convex. Here we first review a key
result in \cite{Palomar06}\cite{Palomar03} that shows with $K=1$ and a single
constraint ($M=1$) the solution of the WSMMSE\ problem can be, however, found
efficiently. We then discuss that with multiple constraints ($M>1$), this is not the case, and
a solution of the WSMMSE\ problem even with $K=1$ must be found through some complex
global optimization strategies. One such technique was recently proposed in
\cite{YuLau11} based on a sophisticated gradient approach. At the end of this section we then propose a
computationally and conceptually simpler solution based on a novel result
(Lemma 5), that our numerical result have shown to have excellent performance.  This will be then leveraged in Sec. \ref{Sec_Multiuser} to propose a novel solution for the general multiuser case.

To elaborate, consider a scenario where the noise-plus-interference matrix
$\mathbf{\Omega}_{k}$ (\ref{eqn:Omega}) is fixed and given (i.e., not subject
to optimization). Now, we solve the WSMMSE problem with $K=1$
for specified weight matrices $\mathbf{W}$ and $\mathbf{\Phi}_{m}$. For the rest of this section, we drop the index $k=1$ from all quantities for simplicity of notation. We proceed by solving the problem at hand, first with respect to
$\mathbf{A}$ for fixed $\mathbf{B},$ and then with respect to
$\mathbf{B}$ without loss of optimality. The first optimization, over
$\mathbf{A}$, is easily seen to be a convex problem (without constraints)
whose solution is given by the minimum MSE equalization matrix
\begin{equation}
\mathbf{A}=\left(  \mathbf{H}\mathbf{B}\mathbf{B}%
^{\mathsf{H}}\mathbf{H}^{\mathsf{H}}+\mathbf{\Omega}\right)
^{-1}\mathbf{H}\mathbf{B}. \label{eqn:MMSEreceiverDownlink}%
\end{equation}
Plugging (\ref{eqn:MMSEreceiverDownlink}) in the MSE matrix
(\ref{eqn:MSE-matrix2}). we obtain
\begin{equation}
\mathbf{E}=\left(  \mathbf{I+B}^{\mathsf{H}}\mathbf{H}%
^{\mathsf{H}}\mathbf{\Omega}^{-1}\mathbf{H}\mathbf{B}\right)
^{-1}. \label{eqn:MMSE_3}%
\end{equation}
We now need to optimize over $\mathbf{B}$ the following problem
\begin{equation}%
\begin{array}
[c]{rl}%
\underset{\mathbf{B}}%
{\text{minimize}} & \tr\left\{\mathbf W\left(\mathbf I+\mathbf B^\mathsf H \mathbf H^\mathsf H \mathbf \Omega^{-1}\mathbf H \mathbf B\right)^{-1}\right\}\\
\text{subject to} & \tr\left\{  \mathbf{\Phi}_{m}%
\mathbf{B}\mathbf{B}^{\mathsf{H}}\right\}  \leq P_{m},
m=1,\ldots,M
\end{array}
, \label{eqn:optimization22}%
\end{equation}

Consider first the single-constraint problem, i.e., $M=1$. The global optimal
solution for single-user WSMMSE problem with $M=1$ is given in \cite{Palomar03}\cite{YuLau11}
and reported below. Recall that, according to Definition \ref{def:IFC_GC}, matrix $\mathbf{\Phi}_{1}$ is positive
definite.

\textit{Lemma 3 \cite{Palomar03}:} The optimal solution of the WSMMSE problem with $K=1$ and a single trace constraint ($M=1$) is given by
\begin{equation}\label{eqn:optimalB_k_su}
\mathbf B=\mathbf \Phi_{1}^{-\frac{1}{2}}\mathbf U \mathbf \Sigma,
\end{equation}
where $\mathbf U \in \mathbb C^{m_{t} \times d}$ 
is the matrix of eigenvectors of matrix $\mathbf \Phi_{1}^{-\frac{1}{2}}\mathbf H \mathbf \Omega^{-1}\mathbf H^\mathsf H\mathbf \Phi_{1}^{-\frac{1}{2}}$ corresponding to its largest eigenvalues $\gamma_{1} \geq \ldots \geq \gamma_{d}$ and $\mathbf \Sigma$ is a diagonal matrix with the diagonal terms $\sqrt{p_{i}}$ defined as
\begin{equation}\label{eqn:optimalp_k_su}
p_{i}=\left[\sqrt{\frac{w_{i}}{\mu\gamma_{i}}}-\frac{1}{\gamma_{i}}\right]^+,
\end{equation}
with $\mu\geq0$ being the ``waterfilling'' level chosen so as to satisfy the
single power constraint $\tr\left\{\mathbf \Phi_{1}\mathbf B \mathbf B^\mathsf H\right\}= P_1$. 

\begin{IEEEproof}
Introducing the ``effective'' precoding matrix $\bar{\mathbf B}=\mathbf \Phi_{1}^{1/2}\mathbf B$ and ``effective'' channel matrix $\bar{\mathbf H}=\mathbf H \mathbf \Phi_{1}^{-\frac{1}{2}}$, the problem is equivalent to the one discussed in \cite[Theorem 1]{Palomar03}.
\end{IEEEproof}

\par In the case of multiple constraints the approach used in Lemma 3 cannot be leveraged.
Here we propose a simple, but effective, approach, which is based on the following considerations
summarized in the following two lemmas.

\par \textit{Lemma 4:} The precoding matrix (\ref{eqn:optimalB_k_su})-(\ref{eqn:optimalp_k_su}) for a given fixed $\mu>0$ minimizes the Lagrangian function
\begin{equation}\label{eqn:Lagrangian_su}
\begin{split}
 \mathcal L(\bar{\mathbf B};\mu)=&\tr\left\{\mathbf W\left(\mathbf I+\bar{\mathbf B}^\mathsf H\mathbf \Phi_{1}^{-\frac{1}{2}}\mathbf H^\mathsf H \mathbf \Omega^{-1} \mathbf H \mathbf \Phi_{1}^{-\frac{1}{2}}\bar{\mathbf B}\right)^{-1}\right\}\\&+\mu\tr\left\{\bar{\mathbf B}\bar{\mathbf B}^\mathsf H\right\}
 \end{split}
 \end{equation}
where $\bar{\mathbf B}$ is the effective precoding matrix defined above.

\begin{IEEEproof} We first note that (\ref{eqn:Lagrangian_su}) is the Lagrangian function of the single-user single-constraint problem solved in Lemma 2. Then, we prove (\ref{eqn:Lagrangian_su}) by contradiction. Assume that the minimum of the Lagrangian function is attained at where the corresponding $\mathbf{E}$ is not diagonal. Then, one can always find a unitary matrix $\mathbf{Q}\in\mathbb{C}^{d\times d}$ such that the matrix $\bar{\mathbf{B}}^\ast=\bar{\mathbf{B}}\mathbf{Q}%
$ diagonalizes $\mathbf{E}$ since with $\bar{\mathbf{B}}^{\ast}$ we
have $\mathbf{E}=\mathbf{Q}^{\mathsf{H}}\left(  \mathbf{I+\bar{B}}%
^{\mathsf{H}}\mathbf \Phi_1^{-\frac{1}{2}}\mathbf{H}^{\mathsf{H}}\mathbf{\Omega}%
^{-1}\mathbf{H}\mathbf \Phi_1^{-\frac{1}{2}}\bar{\mathbf{B}}\right)^{-1}\mathbf{Q}$ \cite{Palomar03}.
The function $\tr\left\{\mathbf W \mathbf E\right\}$ is Schur concave, and therefore the matrix
$\mathbf{B}^{\ast}$ does not decrease the function $\tr\left\{
\mathbf{W}\mathbf{E}\right\}  $ with respect to $\bar{\mathbf B}$, while $\bar{\mathbf B}\bar{\mathbf B}^{\mathsf H}=\bar{\mathbf{B}}^{\ast}\bar{\mathbf{B}}^{\ast\mathsf H}$. This implies that the minimum of $\tr\left\{\mathbf W\mathbf E\right\}$ is reached when the MSE matrix is diagonalized. Therefore, we can set without loss of generality $\bar{\mathbf B}=\mathbf U\mathbf \Sigma$ where $\mathbf U$ is defined as in Lemma 3 and $\mathbf \Sigma$ is diagonal with non-negative elements on the main diagonal. Substituting this form of $\bar{\mathbf B}$ into the Lagrangian function, we obtain a convex problem in the diagonal elements of $\mathbf \Sigma$, whose solution can be easily shown to be given by (\ref{eqn:optimalp_k_su}) for the given $\mu$. This concludes the proof.
\end{IEEEproof}

\textit{Lemma 5:} Let $p^{\star}$ be the optimal
value of the single-user WSMMSE problem with multiple constraints (\mbox{$K=1, M \geq 1$}). We have
\begin{equation}
p^{\star}\geq  \max_{\boldsymbol{\lambda} \geq 0}\inf_{\mathbf B}\mathcal L(\mathbf B;\boldsymbol{\lambda}),\label{eqn:LowerBound}%
\end{equation}
where
\begin{equation}\label{eqn:lagrangian_mc}
\begin{split}
\mathcal L(\mathbf B;\boldsymbol \lambda)=&\tr\left\{\mathbf W\left(\mathbf I+\mathbf B^\mathsf H \mathbf H^\mathsf H \mathbf \Omega\mathbf H\mathbf B\right)^{-1}\right\}\\&+\sum_{m=1}^M\lambda_m\left(\tr\left\{\mathbf \Phi_m\mathbf B\mathbf B^\mathsf H\right\}-P_m\right)
\end{split}
\end{equation}
is the Lagrangian function of the single-user WSMMSE problem at hand and $\boldsymbol \lambda=(\lambda_1,\ldots,\lambda_M)$.
Moreover,
if there exists an optimal solution $\tilde{\mathbf B}$ achieving $p^\star$ that, together with a strictly positive Lagrange multiplier $\tilde{\boldsymbol \lambda} >0$,  satisfies the conditions
 \begin{align}
\nabla_{\mathbf B}\mathcal{L}=0,%
\label{eqn:KKTcond1}\\
\tr\left\{  \mathbf{\Phi}_{m}\tilde{\mathbf{B}}\tilde{\mathbf{B}}%
^{\mathsf{H}}\right\}   &  = P_{m},\quad \forall m\label{eqn:KKTcond2}
\end{align}
then (\ref{eqn:LowerBound}) holds with equality.
\begin{IEEEproof}
The proof is given in the appendix.
\end{IEEEproof}

Lemma 5 suggests that to solve the single-user multiple-constraint problem, under some technical conditions, one can minimize instead the dual problem on the right-hand side of (\ref{eqn:LowerBound}). Lemma 3 showed that this is always possible with a single constraint. The conditions in Lemma 5 hold in most cases where the power constraints for the optimal solution are satisfied with equality. While this may not be always the case, in practice, e.g., if the power constraints represent per-BS power constraints as in the original formulation of Sec. II, this  condition can be shown to hold \cite{Huh10a}.
\par Inspired by Lemma 5, here we propose an iterative approach to the
solution of the WSMMSE problem with $K=1$ that is based on solving the dual problem
$\max_{\boldsymbol{\lambda} \geq 0}\min_{\mathbf B}\mathcal L(\mathbf B;\boldsymbol \lambda).$ Specifically, in
order to maximize $\inf_{\mathbf B}\mathcal L(\mathbf B;\boldsymbol \lambda)$ over $\boldsymbol{\lambda}%
\succeq0$, in the proposed algorithm, the auxiliary variables
$\boldsymbol{\lambda}$ is updated at the $j$th iteration via a subgradient update given by \cite{Liu10a}
\begin{equation}
\lambda_{m}^{(j)}=\lambda_{m}^{(j-1)}+\delta\left(P_m-\tr\left\{
\mathbf{\Phi}_{m}\mathbf{B}\mathbf{B}^{\mathsf{H}}\right\}%
\right),\quad \forall m,
\label{eqn:lambdaUpdate}
\end{equation}
so as to attempt to satisfy the power constraints. Having fixed the vector
$\boldsymbol{\lambda}^{(j)},$ problem $\min_{\mathbf B}\mathcal L(\mathbf B,\boldsymbol \lambda)$ reduces to minimizing (\ref{eqn:Lagrangian_su}) with $\mathbf \Phi_{1}=\mathbf \Phi(\boldsymbol \lambda^{(j)})=\sum_m\lambda_m^{(j)} \mathbf \Phi_{m}$ and $\mu=1$. This can be done using Lemma 3, so that from (\ref{eqn:optimalB_k_su})-(\ref{eqn:optimalp_k_su}), at the $j$th iteration, $\mathbf B^{(j)}$ is obtained as $\mathbf \Phi(\boldsymbol \lambda^{(j)})^{-\frac{1}{2}}\mathbf U^{(j)}\mathbf \Sigma^{(j)}$ where $\mathbf U^{(j)}$ 
is the matrix of eigenvectors of matrix $\mathbf \Phi(\boldsymbol \lambda^{(j)})^{-\frac{1}{2}}\mathbf H \mathbf H^\mathsf H\mathbf \Phi(\boldsymbol \lambda^{(j)})^{-\frac{1}{2}}$ corresponding to its largest eigenvalues $\gamma_{1} \geq \ldots \geq \gamma_{d}$ and $\mathbf \Sigma^{(j)}$ is a diagonal matrix with the diagonal terms $\sqrt{p_{i}}=\sqrt{\left[\sqrt{\frac{w_{i}}{\gamma_{i}}}-\frac{1}{\gamma_{i}}\right]^+}$.

\section{Sum-Rate Maximization}\label{Sec_SRM}
\label{Sec:PreviousTechniques}
The SRM problem for a number of users $K>1$ is non-convex even when removing the constraints on the number of streams per user.
The general problem in fact remains non-convex and is NP-hard
\cite{Luo08}. Therefore, since finding the global optimal has prohibitive complexity, one needs to
resort to suboptimal solutions with reasonable complexity. In this section, we
review several suboptimal solutions to the SRM problem that have been proposed
in the literature. Since some of these techniques were originally proposed for
a scenario that does not subsume the considered MIMO-IFC-GC, we also propose
the necessary modifications required for application to the MIMO-IFC-GC.
Note that these techniques perform an optimization over the transmit covariance matrices by relaxing the rank constraint due to the number of users per streams (see discussion below). Therefore, we also review and modify when necessary a different class of algorithms that solve problems related to SRM but are able to enforce constraints on the number of transmitting streams per user. The WSMMSE problem does not seem to have been addressed previously for the
MIMO-IFC-GC and will be studied in the next section.

\subsection{Soft Interference Nulling\label{Sec:SoftInterferenceNulling}}

A solution to the SRM\ problem for the MIMO-IFC-GC was proposed in \cite{NgHuang10}. In this technique the optimization is over all transmit covariance matrices $\mathbf \Sigma_k=\mathbf{B}_{k}\mathbf{B}_{k}^{\mathsf{H}} \in \mathbb C^{m_{t,k} \times m_{t,k}}$. The constraints on the number of streams would impose a rank constraint on $\mathbf \Sigma_k$ as $\text{rank}(\mathbf \Sigma_k)= d_k$. Here, and in all the following reviewed techniques below, unless stated otherwise, such rank constraints are relaxed by assuming that the number of transmitting data streams is equal to the transmitting antennas to that user, i.e. $d_k=m_{t,k}$. From (\ref{eqn:RateFunction}) and (\ref{eqn:MMSEreceiverDownlink}), we can rewrite the (negative) sum-rate as
\begin{align}
\sum\limits_{k=1}^{K}\log|\mathbf{E}_{k}|    =&-\sum\limits_{k=1}^{K}\log|\mathbf{\Omega
}_{k}+\mathbf{H}_{k,k}\mathbf{\Sigma}_{k}\mathbf{H}_{k,k}^{\mathsf{H}%
}|\nonumber\\
&  +\log|\mathbf{\Omega}_{k}|, \label{eqn:soft}%
\end{align}
where $\mathbf{\Omega}_{k}$ is defined in (\ref{eqn:Omega}).
Notice that it is often convenient to work with the covariance matrices instead of the beamforming matrices $\mathbf{B}_{k},$ since
this change of variables may render the optimization problem convex as, for
instance, when minimizing the first term only in (\ref{eqn:soft}). It can then
be seen that the SRM\ problem is, however, non-convex due to the presence of
the $-\log|\mathbf{\Omega}_{k}|$ term, which is indeed a concave function of
the matrices $\mathbf{\Sigma}_{k}.$

An approximate solution is then be found in \cite{NgHuang10} via an iterative scheme, whereby at
each $(j+1)$th iteration, given the previous solution $\mathbf{\Sigma}_{k}^{(j)}$
the non-convex term $-\log|\mathbf{\Omega}_{k}|$ is approximated using a
first-order Taylor expansion as
\begin{equation}
\begin{split}
-\log|\mathbf{\Omega}_{k}|\simeq & -\log|\mathbf{\Omega}^{(j)}_{k}|\\&-\sum\limits_{l \neq k}\tr\left\{\left(\mathbf \Omega_k^{(j)}\right)^{-1}\mathbf H_{k,l}\left(\mathbf \Sigma_l-\mathbf \Sigma_l^{(j)}\right)\mathbf H_{k,l}^\mathsf H\right\}, \label{eqn:taylor}
\end{split}
\end{equation} where
$\mathbf{\Omega}_{k}^{(j)}=\mathbf{I}+\sum\limits_{l\neq k}\mathbf{H}_{k,l}\mathbf{\Sigma}_k^{(j)}\mathbf{H}_{k,l}^{\mathsf{H}}$. Since the resulting problem

\begin{equation}%
\begin{array}
[c]{rl}%
\underset{\mathbf \Sigma_k, k=1,\ldots,K}{\text{minimize}} & -\sum\limits_{k=1}^K \log\left|\mathbf \Omega_k+\mathbf H_{k,k}\mathbf \Sigma_k\mathbf H_{k,k}\right|\\
& +\sum\limits_{l \neq k}\tr\left\{\left(\mathbf \Omega_k^{(j)}\right)^{-1}\mathbf H_{k,l}\mathbf \Sigma_l\mathbf H_{k,l}^\mathsf H\right\}\\
\text{subject to} & \tr\left\{  \mathbf{\Phi}_{k,m}\mathbf \Sigma_k\right\}  \leq P_{m},\quad
m=1,\ldots,M,
\end{array}
\label{eqn:soft_opt}%
\end{equation}
is convex, a solution can be found efficiently. Following the original
reference \cite{NgHuang10}, we refer to this scheme as ``soft interference nulling''.
We refer to \cite{NgHuang10} for further details about the algorithm.

\subsection{SDP Relaxation}\label{Sec_SDPRelaxation}
A related approach is taken in \cite{Luo10} for the
SRM\ problem\footnote{More generally, the reference studies the weighted
SRM\ problem.} for a MIMO-IFC with regular per-transmitter, rather than
generalized, power constraints. Similarly to the previous technique, the optimization is over the transmit covariance matrices and under the relaxed rank constraints. In particular, the authors first approximate
the problem by using the approach in \cite{Christensen08}. Then, an
iterative solution is proposed by linearizing a non-convex term similar to
soft interference nulling as reviewed above. It turns out that such linearized
problem can be solved using semi-definite programming (SDP). Specifically, denoting
with $\mathbf{\Omega}_{k}^{(j)}$ the matrix (\ref{eqn:Omega}) corresponding to
the solution $\mathbf{B}_{k}^{(j)}$ at the previous iteration $j$, i.e.,
$\mathbf{\Omega}_{k}^{(j)}=\mathbf{I}+\sum_{l\neq k}\mathbf{H}_{k,l}%
\mathbf{B}_{l}^{(j)}\mathbf{B}_{l}^{(j)\mathsf{H}}\mathbf{H}_{k,l}%
^{\mathsf{H}},$ the SDP problem to be solved to find the solutions
$\mathbf{B}_{k}^{(j+1)}$ for the $(j+1)$th iteration is given by%
\[%
\begin{array}
[l]{ll}%
\begin{array}
[l]{l}%
\underset{\mathbf{Y}_{k},\mathbf{\Sigma}_{k},\forall k}{\text{minimize}}
\end{array}
& \sum\limits_{k=1}^{K}\tr\left\{  \mathbf{Y}_{k}\right\}  +\sum\limits_{k=1}^{K}\tr\left\{
\mathbf{C}_{k}^{(j)}\mathbf{\Sigma}_{k}\right\}  \\
\text{subject to} & \sum\limits_{k=1}^{K}\tr\left\{  \mathbf{\Phi}_{k,m}%
\mathbf{\Sigma}_{k}\right\}  \leq P_{m},\quad m=1,\ldots,M\\
& \left[\!\!
\begin{array}
[c]{cc}%
\mathbf{H}_{k,k}\mathbf{\Sigma}_{k}\mathbf{H}_{k,k}^{\mathsf{H}}%
+\mathbf{\Omega}_{k}^{(j)}\! & \!\left(  \mathbf{W}_{k}^{(j)}\mathbf{\Omega}%
_{k}^{(j)}\right)  ^{\frac{1}{2}}\\
\left(  \mathbf{W}_{k}^{(j)}\mathbf{\Omega}_{k}^{(j)}\right)  ^{\frac{1}{2}} \!&\!
\mathbf{Y}_{k}%
\end{array}\!\!
\right]  \succeq0,\\
& \text{and }\mathbf{\Sigma}_{k}\succeq0,\quad k=1,\ldots,K
\end{array}
\]
where%
\begin{equation}
\mathbf{W}_{k}^{(j)}=\mathbf{I}+\mathbf{H}_{k,k}\mathbf{\Sigma}_{k}%
^{(j)}\mathbf{H}_{k,k}^{\mathsf{H}}\mathbf{\Omega}_{k}^{(j)-1},%
\end{equation}
\begin{align}
\mathbf{C}_{k}^{(j)} &  =\sum\limits_{i\neq k}\mathbf{H}_{i,k}^{\mathsf{H}}\left(
\mathbf I+ \sum\limits_{l}\mathbf{H}_{i,l}\mathbf{\Sigma}_{l}^{(j)}%
\mathbf{H}_{i,l}^{\mathsf{H}}\right)  ^{-1}\mathbf{W}_{i}^{(j)}\times
\nonumber\\
&  \mathbf{H}_{i}\mathbf{\Sigma}_{i}^{(j)}\mathbf{H}_{i}^{\mathsf{H}}\left(
\mathbf I+ \sum\limits_{l}\mathbf{H}_{i,l}\mathbf{\Sigma}_{l}^{(j)}%
\mathbf{H}_{i,l}^{\mathsf{H}}\right)  ^{-1}\mathbf{H}_{i,k},
\end{align}
and $\mathbf Y_k$ is an auxiliary optimization variable, defined using the Schur complement as $\mathbf Y_k=\mathbf W_k\mathbf \Omega_k^{(j)}\left(\mathbf H_{k,k}\mathbf \Sigma_k\mathbf H_{k,k}^\mathsf H+\mathbf \Omega_k^{(j)}\right)^{-1}$ to convert the original optimization problem to an SDP problem \cite{Luo10}.
The derivation requires minor modifications with respect to \cite{Luo10} and
is therefore not detailed. The scheme is referred to as ``SDP\ relaxation'' in
the following. We refer to \cite{Luo10} for further details about the algorithm.

\subsection{Polite Waterfilling}\label{Sec_PoliteWaterFilling}
\label{Sec:PoliteWaterFilling}Reference \cite{Liu10a} studied the (weighted)
SRM\ problem for a general model that includes the MIMO-IFC-GC. We review the approach here for completeness. Two algorithms
are proposed, whose basic idea is to search iteratively for a solution of the
KKT\ conditions \cite{Boyd04} for the (weighted) SRM problem. Notice that, since the
problem is non-convex, being a solution of the KKT conditions is only
necessary (as proved in \cite{Liu10a}) but not sufficient to guarantee global
optimality. It is shown in \cite{Liu10a} that any solution $\mathbf{\Sigma
}_{k},$ $k=1,\ldots,K$, of the KKT conditions must have a specific structure that
is referred to as ``polite waterfilling'', which is reviewed below for the SRM problem.

\textit{Lemma 6 \cite{Liu10a}:} For a given set of Lagrange multipliers
$\boldsymbol{\lambda=}(\mu\lambda_{1},...,\mu\lambda_{M}),$ where $\mu>0$ and
$\lambda_{i}\geq0$ for $i=1,...,M$, associated with the $M$ power constraints
in (\ref{eqn:optimization11}), define the covariance matrices
\begin{equation}
\mathbf{\hat{\Omega}}_{k}=\sum\limits_{m=1}^{M}\lambda_{m}\mathbf{\Phi}%
_{k,m}+\sum\limits_{j\neq k}\mathbf{H}_{j,k}^{\mathsf{H}}\mathbf{\hat{\Sigma}}_{j}\mathbf{H}_{j,k},\label{eqn:omega_hat}%
\end{equation}
with
\begin{equation}
\mathbf{\hat{\Sigma}}_{k}=\frac{1}{\mu}\left(  \mathbf{\Omega}_{k}%
^{-1}-\left(  \mathbf{\Omega}_{k}+\mathbf{H}_{k,k}\mathbf{\Sigma}%
_{k}\mathbf{H}_{k,k}^{\mathsf{H}}\right)  ^{-1}\right)  .\label{eqn:Sigma_hat}%
\end{equation}
An optimal solution $\mathbf{\Sigma}_{k},$ $k=1,...,K$, of the SRM\ problem
must have the ``polite waterfilling'' form
\begin{equation}
\mathbf{\Sigma}_{k}=\mathbf{\hat{\Omega}}_{k}^{-\frac{1}{2}}\mathbf{V}%
_{k}\mathbf{P}_{k}\mathbf{V}_{k}^{\mathsf{H}}\mathbf{\hat{\Omega}}%
_{k}^{-\frac{1}{2}},\label{eqn:pp}%
\end{equation}
where the columns of $\mathbf{V}_{k}$ are the right singular vectors of the ``pre- and post- whitened channel matrix'' $\mathbf{\Omega}_{k}^{-\frac{1}{2}}%
\mathbf{H}_{k,k}\mathbf{\hat{\Omega}}_{k}^{-\frac{1}{2}}$ with
(\ref{eqn:Omega}) for $k=1,\ldots,K$, and $\mathbf{P}_{k}$ is a diagonal matrix
with diagonal elements $p_{k,i}.$ The powers $p_{k,i}$ must satisfy
\begin{equation}
p_{k,i}=\left[  \frac{1}{\mu}-\frac{1}{\gamma_{k,i}}\right]  ^{+}%
,\label{eqn:optimalDeltaPolite}%
\end{equation}
where $\sqrt{\gamma_{k,i}}$ is the $i$th singular value of the whitened matrix
$\mathbf{\Omega}_{k}^{-\frac{1}{2}}\mathbf{H}_{k,k}\mathbf{\hat{\Omega}}%
_{k}^{-\frac{1}{2}}.$ Parameter $\mu\geq0$ is selected so as to satisfy the
constraint%
\begin{equation}
\sum\limits_{m=1}^{M}\lambda_{m}\sum\limits_{k=1}^{K}\tr\left\{  \mathbf{\Phi}%
_{k,m}\mathbf{\Sigma}_{k}\right\}  \leq\sum\limits_{m=1}^{M}\lambda_{m}%
P_{m},\label{eqn:sumconst}%
\end{equation}
which implied by the constraints of the original problem
(\ref{eqn:optimization11}). Moreover, parameters $\lambda_{i}\geq0$ are to be
chosen so as to satisfy each individual constraint in
(\ref{eqn:optimization11}).

In order to obtain a solution $\mathbf{\Sigma}_{k},$ $k=1,\ldots,K$, according to polite waterfilling form as described in Lemma 6, \cite{Liu10a} proposes to use the interpretation of $\hat{\mathbf \Omega}_k$ in (\ref{eqn:omega_hat}) as the interference plus noise covariance matrix and $\hat{\mathbf \Sigma}_k$ in (\ref{eqn:Sigma_hat}) as the transmit covariance matrix both at the ``dual'' system\footnote{In the ``dual'' system the role of transmitters and receivers is
switched, i.e., the $k$th transmitter in the original system becomes the $k$th
receiver in the ``dual'' system. The channel matrix between the $k$th
transmitter and the $l$th receiver in the dual system is given by
$\mathbf{H}_{l,k}^{\mathsf{H}}.$}.

Based on this observation, the algorithm proposed in \cite{Liu10a} works as
follows. At each $j$th iteration, first one calculates the covariance matrices
$\mathbf{\Sigma}_{k}^{(j)}$ in the original system using the polite
waterfilling solution of Lemma 6; then one calculates the matrices
$\mathbf{\hat{\Sigma}}_{k}^{(j)}$ using again polite waterfilling in the dual system as
explained above. Finally, at the end of each $j$th iteration, one updates the
Lagrange multipliers as
\begin{equation}
\lambda_{m}^{(j+1)}=\lambda_{m}^{(j)}\frac{\sum\limits_{k=1}^{K}\tr\left\{
\mathbf{\Phi}_{k,m}\mathbf{\Sigma}_{k}^{(j)}\right\}  }{P_{m}},
\end{equation}
thus forcing the solution to satisfy the constraints of the SRM problem
(\ref{eqn:optimization11}). For details on the algorithm, we refer to \cite{Liu10a}.

\begin{remark} Other notable algorithms designed to solve the SRM problem
for the special case of a MIMO-BC with generalized constraints are \cite{Huh09,Huh10}. As explained in \cite{Liu10a}, these schemes are not easily
generalized to the scenario at hand where the cost function is not convex. As
such, they will not be further studied here.
\end{remark}
\subsection{Leakage Minimization}
While the techniques discussed above do not enforce constraints on the number of stream per users, here we extend a technique previously proposed in \cite{Gomadam08} that aims at aligning interference through minimizing the interference leakage and is able to enforce the desired rank constraints. It is known that this approach is solves the SRM problem for high signal-to-noise-ratio (SNR). In this algorithm, it is assumed that the power budget is divided equally between the data streams and the precoding matrix of user $k$ from BS $m$ is given as $\mathbf B_{k,m}=\sqrt{\frac{P_m}{K_md_k}}\bar{\mathbf B}_{k,m}$ where $\bar{\mathbf B}_{k,m}$ is a $n_t \times d_k$ matrix of orthonormal columns (i.e. $\bar{\mathbf B}_{k,m}^\mathsf H \bar{\mathbf B}_{k,m}=\mathbf I$). The equalization matrices are also assumed to have orthonormal columns (i.e. $\mathbf A_k^\mathsf H \mathbf A_k=\mathbf I$). Hence, there is no inter-stream interference for each user. Total interference leakage at user $k$ is given by
\begin{equation}
I=\sum_k\tr\left\{\mathbf A_k^\mathsf H \mathbf Q_k \mathbf A_k\right\}.
\label{eqn:Leakage}
\end{equation}
where $\mathbf Q_k=\sum_{j\neq k}\sum_{m \in \mathcal M_j}\frac{P_m}{K_md_j}\widetilde{\mathbf H}_{k,m}\bar{\mathbf B}_{k,m}\bar{\mathbf B}_{k,m}^\mathsf H\widetilde{\mathbf H}_{j,m}^\mathsf H$. To minimize the interference leakage, the equalization matrix $\mathbf A_k$ for user $k$ can be obtained as $\mathbf A_k=v_{d_k}(\mathbf Q_k)$ where $v_{d_k}(\mathbf A)$ represents a matrix with columns given by the eigenvectors corresponding to the $d_k$ smallest eigenvalues of $\mathbf A$. Now, for fixed matrices $\mathbf A_k$, the cost function (\ref{eqn:Leakage}) can be rewritten as
\begin{equation}
I=\sum_k\sum_{m \in \mathcal M_k}\tr\left\{\bar{\mathbf B}_{k,m}^\mathsf H \hat{\mathbf Q}_{k,m}\bar{\mathbf B}_{k,m}\right\}
\end{equation}
where $\hat{\mathbf Q}_{k,m}=\sum_{j \neq k, j \in \mathcal K_m}\frac{P_m}{K_md_k}\widetilde{\mathbf H}_{j,m}^\mathsf H\mathbf A_j\mathbf A_j^\mathsf H \widetilde{\mathbf H}_{j,m}$.\footnote{In the original work \cite{Gomadam08} which is proposed for the interference channels, the algorithm iteratively exchanges the role of transmitters and receivers to update the precoding and equalization matrices similarly.} Minimizing over the matrices $\mathbf B_k$ leads to choosing $\bar{\mathbf B}_{k,m}=v_{d_k}(\hat{\mathbf Q}_{k,m})$. The algorithm iterates until convergence. We refer to this scheme as ``min leakage'' in the following.

\subsection{Max-SINR}
Another algorithm called ``max-SINR'' has been proposed in \cite{Gomadam08} which is based on the maximization of SINR, rather than directly the sum-rate. This algorithm is also able to enforce rank constraints. The max-SINR algorithm assumes equal power allocated to the data streams and attempts at maximizing the SINR for each stream by selecting the receive filters. Then, it exchanges the role of transmitter and receiver to obtain the transmit precoding matrices which maximizes the max-SINR. This iterates until convergence. A modification of this algorithm is given in \cite{Peters10} by maximizing the ratio of the average signal power to the interference plus noise power (SINR-like) term. However, these techniques are only given for standard MIMO interference channels and not for MIMO-IFC-GC.

\section{MSE Minimization}\label{Sec_MSEMinimization}

In this section, we propose two suboptimal techniques to solve the
WSMMSE problem. We recall that with the WSMMSE problem enforcing the constraint on $d_k$ is necessary in order to avoid trivial solutions. Performance comparison among all the considered schemes will be
provided in Sec.~\ref{sec_NumericalResults} for a multi-cell system with network MIMO.

\subsection{MMSE Interference Alignment}

A technique referred to as MMSE interference alignment (MMSE-IA) was presented
in \cite{Schmidth09} for an interference channel with per-transmitter power
constraints and where each receiver is endowed with a single antenna. Here we
extend the approach to to the MIMO-IFC-GC.

The idea is to approximate the solution of the WSMMSE problem by
optimizing the precoding matrices $\mathbf{B}_{k}$ followed by the equalization matrices $\mathbf{A}_{k}$ and iterating the procedure. Specifically, initialize $\mathbf{B}_{k}$
arbitrarily. Then, at each iteration $j$: (\textit{i}) For each user $k$,
evaluate the equalization matrices using the MMSE solution
(\ref{eqn:MMSEreceiverDownlink}), obtaining $\mathbf{A}_{k}^{(j)}\!=\!\left(\!
\mathbf{H}_{k,k}\mathbf{B}_{k}^{(j-1)}\mathbf{B}_{k}^{(j-1)\mathsf{H}%
}\mathbf{H}_{k,k}^{\mathsf{H}}\!+\!\mathbf{\Omega}_{k}^{(j-1)}\right)
^{-1}\mathbf{H}_{k,k}\mathbf{B}_{k}^{(j-1)},$ where from (\ref{eqn:Omega}) we
have $\mathbf{\Omega}_{k}^{(j-1)}\!=\!\mathbf{I}\!+\!\sum_{l\neq k}\!\mathbf{H}%
_{k,l}\mathbf{B}_{l}^{(j-1)}\mathbf{B}_{l}^{(j-1)\mathsf{H}}\mathbf{H}%
_{k,l}^{\mathsf{H}}$; (\textit{ii}) Given the matrices $\mathbf{A}_{k}^{(j)}$,
the WSMMSE problem becomes%
\begin{equation}%
\begin{array}
[c]{rl}%
\underset{\mathbf{B}_{k}\text{,\textbf{ }}k=1,...,K}{\text{minimize}} &
\sum\limits_{k=1}^{K}\tr\left\{\mathbf W_k\mathbf E_k^{(j)}\right\}\\
\text{subject to} & \sum\limits_{k=1}^{K}\tr\left\{  \mathbf{\Phi}_{k,m}%
\mathbf{B}_{k}\mathbf{B}_{k}^{\mathsf{H}}\right\}  \leq P_{m},\forall m \in \mathcal M
\end{array}
,\label{eqn:eMMSE}%
\end{equation}
where $\mathbf{E}_{k}^{(j)}$ is (\ref{eqn:MSE-matrix2}) with $\mathbf{A}%
_{k}^{(j)}$ in place of $\mathbf{A}_{k}.$ Fixing the equalization matrices $\mathbf A_k^{(j)}, \forall k$, this problem is convex in
$\mathbf{B}_{k}$ and can be solved by enforcing the KKT\ conditions.
Therefore, matrices $\mathbf{B}_{k}^{(j)}$ for the $j$th iteration can be
obtained as follows.

\textit{Lemma 7:} For given equalization matrices $\mathbf{A}_{k}^{(j)}$, a solution
$\mathbf{B}_{k}^{(j)}$,\textbf{ }$k=1,...,K,$ of the WSMMSE problem is given by
\begin{equation}\label{eqn:Bk_MMSEIA}
\begin{split}
\mathbf B_k^{(j)}=\left(\sum\limits_{l=1}^K \mathbf H_{l,k}^\mathsf H \mathbf A_l^{(j)} \mathbf W_l\mathbf A_l^{(j)\mathsf H}\mathbf H_{l,k}+\sum\limits_m\mu_m\mathbf \Phi_{k,m}\right)^{-1}\times\\ \mathbf H_{k,k}^\mathsf H \mathbf A_k^{(j)} \mathbf W_k
\end{split}
\end{equation}
where $\mu_m$ are Lagrangian multipliers satisfying
\begin{align}
\mu_{m}  &  \geq 0\label{eqn:kkt6}\\
\mu_{m}\left(  \sum\limits_{k=1}^K\tr\left\{\mathbf \Phi_{k,m}\mathbf B_k^{(j)}\mathbf B_k^{(j)\mathsf H}\right\}-P_{m}\right)   &  =0
\label{eqn:kkt7}%
\end{align}
and the power constraints $\sum_{k=1}^K\tr\left\{\mathbf \Phi_{k,m}\mathbf B_k^{(j)}\mathbf B_k^{(j)\mathsf H}\right\}\leq P_{m}$ for all $m$.
\par Once obtained the matrices $\mathbf{B}_{k}^{(j)}$ using the results in
Lemma 7, the iterative procedure continues with the ($j+1$)th iteration.
We refer to this scheme as extended MMSE-IA, or eMMSE-IA.

\begin{remark}
The algorithm proposed above reduces to the one introduced in
\cite{Schmidth09} in the special case of per-transmitter power constraints and
single-antenna receivers. It is noted that in such case, problem
(\ref{eqn:eMMSE}) can be solved in a distributed fashion, so that each
transmitter $k$ can calculate its matrix (more precisely vector, given the
single antenna at the receivers) independently from the other transmitters. In
the MIMO-IFC-GC, the power constraints couple the solutions of the different users
and thus make a distributed approach infeasible.
\end{remark}

\subsection{Diagonalized MMSE}\label{Sec_Multiuser}

We now propose an iterative optimization strategy inspired by the single-user
algorithm that we put forth in Sec. \ref{sec_SingleUser}. At the ($j+1)$th iteration, given the
matrices obtained at the previous iteration, we proceed as follows. The
weighted sum-MSE (\ref{eqn:WeightedMSEfunction}) with the definition of MSE-matrices (\ref{eqn:MSE-matrix2}) is a convex function in each $\mathbf A_k$ and $\mathbf B_k$ when $(\mathbf B_j, \mathbf A_j), \forall j \neq k$ are fixed. Nevertheless, it is not jointly convex in terms of both $(\mathbf A_k, \mathbf B_k)$. Inspired by Lemma 5 for the corresponding single-user problem, we propose a (suboptimal) solution based on the solution of the dual problem for calculation of $(\mathbf A_k,\mathbf B_k)$. To this end, we first obtain $\mathbf A_k$ as (\ref{eqn:MMSEreceiverDownlink}). Then, we simplify the Lagrangian function with respect to $\mathbf B_k$ by removing the terms independent of $\mathbf B_k$. Specifically, by defining $\mathbf{\Upsilon}_{k}%
=\sum_{l\neq k}\mathbf{H}_{l,k}^{\mathsf{H}}\mathbf{A}_{l}\mathbf{W}%
_{l}\mathbf{A}_{l}^\mathsf{H}\mathbf{H}_{l,k}$, we have that the Lagrangian function at hand is given by
\begin{equation}\label{eqn:Lagrangian_mu}
\begin{split}
\mathcal L(\mathbf B_k;\boldsymbol \lambda)=& \tr\left\{\mathbf W_k\left(\mathbf I+\mathbf B_k^\mathsf H \mathbf H_{k,k}^\mathsf H\mathbf \Omega_k^{-1}\mathbf H_{k,k} \mathbf B_k\right)^{-1}\right\} \\& +\tr\left\{\mathbf \Upsilon_k\mathbf B_k \mathbf B_k^\mathsf H\right\}\\&+\tr\left\{\left(\sum\limits_m\lambda_m\mathbf \Phi_{k,m}\right)\mathbf B_k\mathbf B_k^\mathsf H\right\}
\end{split}
\end{equation}
This Lagrangian function for user $k$ is the same as the Lagrangian function (\ref{eqn:lagrangian_mc}) of single-user WSMMSE problem when $\mathbf \Phi(\boldsymbol \lambda)$ is replaced with $\mathbf F_k(\boldsymbol \lambda)=\mathbf \Upsilon_k+\sum\lambda_m\mathbf \Phi_{k,m}$.  Matrix $\mathbf F_k(\boldsymbol \lambda)$ is non-singular and therefore, using the same argument as in the proof of Lemma 5, for a given Lagrange multipliers $\boldsymbol \lambda$ and given other users' transmission strategies $(\mathbf A_l,\mathbf B_l), \forall l \neq k$, the optimal transmit precoding matrix can be obtained as
\begin{align}
\mathbf B_k=\mathbf F_{k}(\boldsymbol \lambda)^{-\frac{1}{2}}\mathbf U_k \mathbf \Sigma_k, \label{eqn:optimalB_k_mu}
\end{align}
where $\mathbf U_k \in \mathbb C^{m_{t,k} \times d_k}$ is the eigenvectors of $\mathbf F_{k}(\boldsymbol \lambda)^{-\frac{1}{2}}\mathbf H_{k,k}^\mathsf H \mathbf \Omega_k^{-1}\mathbf H_{k,k}\mathbf F_{k}(\boldsymbol \lambda)^{-\frac{1}{2}}$ corresponding to its largest eigenvalues $\gamma_{k,1}\geq \ldots \geq \gamma_{k,d_k}$ and $\mathbf \Sigma_k$  is diagonal matrices with the elements $\sqrt{p_{k,i}}$ given by
\begin{align}
p_{k,i}=\left[\sqrt{\frac{w_{k,i}}{\gamma_{k,i}}}-\frac{1}{\gamma_{k,i}}\right]^+,
\label{eqn:optimalp_k_mu}
\end{align}
with $\boldsymbol \lambda \succeq 0$ being the Lagrangian multipliers satisfy the power constraints.
Since this scheme diagonalizes the MSE matrices defined in (\ref{eqn:MSEdefined}), it is referred to as diagonalized MMSE (DMMSE).

To summarize, the proposed algorithm at each iteration $j$ (\textit{i}) evaluates the transmit precoding matrices $\mathbf B_k^{(j)}$ given other users' transmission strategies $(\mathbf A_l^{(j-1)},\mathbf B_l^{(j-1)})$ using (\ref{eqn:optimalB_k_mu})-(\ref{eqn:optimalp_k_mu}) (\textit{ii}) updates the equalization matrices using the MMSE solution (\ref{eqn:MMSEreceiverDownlink}); (\textit{iii}) updates the $\boldsymbol \lambda$ via a subgradient update
\begin{equation}
\lambda_m^{(j+1)}=\lambda_m^{(j)}+\delta\left(P_m-\sum\limits_{k=1}^K\tr\left\{\mathbf \Phi_{k,m}\mathbf B_k\mathbf B_k^\mathsf H\right\}\right)
\end{equation}
 to satisfy the power constraints.
\begin{remark}
In this paper, we assume perfect knowledge of channel state information (CSI). Therefore, each transmitter and receiver has sufficient information to calculate the resulting precoders and equalizers by running the proposed algorithms. Under this assumption, which is common to other reviewed works such as \cite{NgHuang10}\cite{Luo10}, no exchange of precoder and equalizer vectors is required between the transmitters and receivers. Nevertheless, in practice, the CSI may only be available locally, in the sense that transmitter $k$ knows channel matrices $\mathbf H_{l,k}$, for all $l=1,\ldots,K$, whereas receiver $k$ is aware of channel matrices $\mathbf H_{k,l}$, for all $l=1,\ldots,K$. The proposed DMMSE and the reviewed PWF \cite{Liu10}\cite{Liu10a} algorithms require, beside the local CSI, that the transmitter $k$ has available also the interference plus noise covariance matrix, $\mathbf \Omega_k$, and the current equalization matrices $\mathbf A_l$ for all $l=1,\ldots,K$ in order  to update the precoder for user $k$. Hence, to enable DMMSE and PWF with local CSI, exchange of the equalizer matrices is needed between the nodes. Similarly, the proposed eMMSEIA, and min leakage and Max-SINR algorithms \cite{Gomadam08}, require the transmitters to know the equalizing matrices $\mathbf A_l$ for $l=1,\ldots,K$ at each iteration, in addition to the local CSI. Moreover, each receiver must know the current precoders $\mathbf B_l$ for all $l=1,\ldots,K$. Therefore, the overhead for the proposed eMMSEIA and the min leakage and Max-SINR algorithms involves the exchange of equalizer and precoder matrices between the transmitters and receivers. However, these latter algorithms can also be adapted using the bi-directional optimization process proposed in \cite{Shi10}. This process involves bi-directional training followed by data transmission. In the forward direction, the training sequences are sent using the current precoders. Then, at the user receivers the equalizers are updated to minimize the least square error cost function. In the backward training phase, the current equalizers are used to send the training sequences and the precoders are updated accordingly. Finally, the SIN \cite{NgHuang10} and SDP relaxation \cite{Luo10} techniques are applied in a centralized fashion (rather than by updating the transmitter and receiver for each user at each iteration), and they require centralized full knowledge of all channel matrices.
\end{remark}
\begin{remark}
 Reference \cite{Luo10} addresses the SRM problem for a MIMO-IFC with regular per-transmitter, rather than generalized, power constraints. The problem is addressed by solving an SDP problem at each iteration. Moreover, the optimization is over the transmit covariance matrices and under the relaxed rank constraint. This enforces a constraint on the number of transmitted streams per user. References \cite{Liu10}-\cite{Liu10a} study the (weighted) SRM problem by decomposing the multiuser problem into single-user problems for each user. Each single-user problem is a standard single-user SRM problem with an additional interference power constraint. The approach used in \cite{Liu10}-\cite{Liu10a} assumes that the number of transmitted streams is equal to $n_r$. In this paper, we address WSMMSE problem and allow for
  an arbitrary number of streams ($d_k \leq  n_r$).
\end{remark}
\begin{remark}
Our algorithms consists of an inner loop, which solves the WSMMSE problem, and an outer loop, which is the subgradient algorithm to update $\boldsymbol \lambda$. The subgradient algorithm in the outer loop is convergent (with a proper selection of the step sizes \cite{Bertsekas03}) due to the fact that the dual function $\text{inf}_\mathbf B\mathcal L(\mathbf B;\boldsymbol \lambda)$ is a concave function with respect to $\boldsymbol \lambda$ \cite{Boyd04}. The inner loops of the proposed algorithms in this paper (i.e. eMMSEIA and DMMSE) are convergent since the objective function decreases at each iteration. A discussion of the convergence for a special case of the eMMSEIA algorithm can be found in \cite{Schmidth09}. However, the original problem is non-convex and our solutions are only local minima. Nevertheless, the DMMSE algorithm is shown to converge to a local minimum with better performance compared to the previously known schemes in Sec.~\ref{sec_NumericalResults}.
\end{remark}
\section{Numerical Results}\label{sec_NumericalResults}
We consider a hexagonal cellular system where each BS is equipped with $n_{t}$ transmit
antennas and each user has $n_{r}$ receive antennas. The users are
located uniformly at random. 
Two tiers of surrounding cells are considered as interference for each cluster. We consider the worst-case scenario for the inter-cluster interference, which will be the condition that interfering BSs transmit at the full allowed power
\cite{Ye05,Zhang09,Huang09,Kaviani10a}. We define the cooperation factor $\kappa$ as
a number of BSs cooperating on transmission to each user. The $\kappa$ BSs are assigned to each user so that the corresponding channel norms (or, alternatively, the corresponding received SNRs) are the largest.

The propagation channel between each BS's transmit antennas and mobile
user's receive antenna is characterized by path loss, shadowing and Rayleigh
fading. The path loss component is proportional to $d_{km}^{-\beta}$, where
$d_{km}$ denotes distance from base station $m$ to mobile user $k$ and
$\beta=3.8$ is the path loss exponent. The channel from the transmit antenna $t$ of the base station $b$ at the receive
antenna $r$ of the $k$th user is given by \cite{Huang09}
\begin{equation}
\mathbf{H}_{k,b}^{(r,t)}=\alpha_{k,b}^{(r,t)}\sqrt{\gamma_{0}\rho_{k,b}A\left(\Theta_{k,b}^{(t)}\right) \left( \frac{d_{k,b}}{d_{0}}\right) ^{-\beta}}%
\end{equation}
where $\alpha_{k,b}^{(r,t)} \sim\mathcal{CN}\left( 0,1\right) $ represents
Rayleigh fading, $\rho_{k,b}^{(\text{dBm})}$ is the lognormal shadow fading
between $b$th BS and $k$th user with standard deviation of $8 \text{ dB}$, and
$d_{0}=1 \text{ km}$ is the cell radius. $\gamma_{0}$ is the
interference-free SNR at the cell boundary. We consider one user randomly located per cell for the numerical results.

\par When sectorization is employed, the
transmit antennas are equally divided among the sectors of a cell.
Each transmit antenna has a parabolic beam pattern as a function of the direction of the user from the broadside direction of the antenna (For more details refer to \cite{Stutzman97,Huang09}). The antenna gain is a function of the direction of the user $k$ from the broadside direction of the $t$th transmit antenna of the $b$th base station denoted by $\Theta_{k,b}^{(t)} \in\left[ -\pi,\pi\right] $; $\Theta_{\mathsf {3dB}}$ is the half-power angle and $A_s$  is the sidelobe gain. The antenna gain is given as
\cite{Stutzman97}
\begin{equation}
\label{eqn:antennapattern}A\left( \Theta_{k,b}^{(t)}\right) _{\mathsf {dB}}=-\min\left(
12\left( \frac{\Theta_{k,b}^{(t)}}{\Theta_{\mathsf{3 dB}}}\right) ^{2}%
,A_{s}\right)
\end{equation}
For the 3,6-sector cells $A_{s}=20,23 \text{ dB}$ and $\Theta
_{\mathsf{3dB}}=\frac{70\pi}{180},\frac{35\pi}{180}$, respectively \cite{Stutzman97,Huang09,Huang10}. When there is no
sectorization we set $A=1$.

\par We first compare different algorithms (for the solution of the SRM problem) without enforcing rank constraints on SIN, PWF, SDP relaxation and setting $d_k=\min(m_{t,k},m_{r,k})=n_r$ for the eMMSEIA and DMMSE algorithms. To solve the SRM problem, the weight matrices in the eMMSEIA and DMMSE algorithms are updated at each iteration as $\mathbf W_k=\mathbf E_k^{-1}$ using the current MSE-matrix $\mathbf E_k$.
\begin{figure}[t!]
\centering
\includegraphics[width=3.8in]{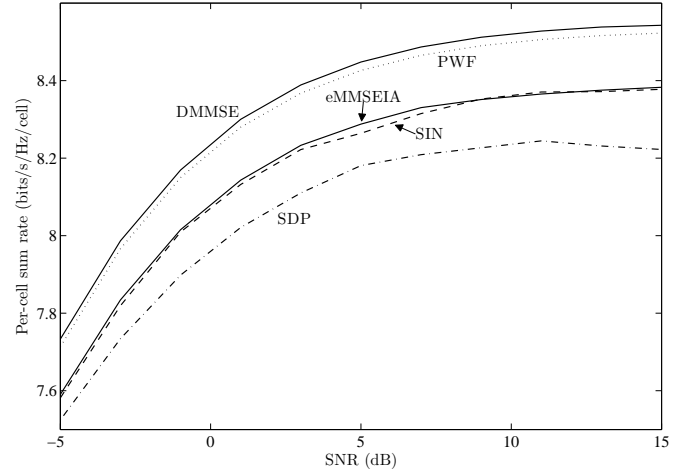}  \caption{Per-cell sum-rate for a MIMO-IFC-GC
with $M=3$ and $\kappa=2$.}%
\label{Fig:2-8}%
\end{figure}
\begin{figure}[t!]
\centering
\includegraphics[width=3.8in]{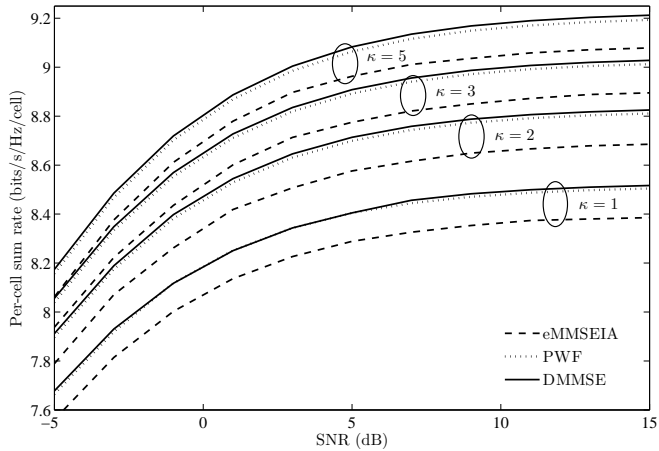}  \caption{Per-cell sum-rate for a MIMO-IFC-GC
with $M=5$ and $\kappa=1,2,3,5$, $n_t=4$, $n_r=d_k=2$, and 2 users per cell.}%
\label{Fig:2-9}%
\end{figure}
Fig.~\ref{Fig:2-8} compares the per-cell sum-rate of the algorithms discussed in
this paper for a cluster with $M=3$ cells and a cooperation factor $\kappa=2$. The results show that our proposed DMMSE algorithm outperforms other techniques, while the polite water-filling algorithm (PWF) \cite{Liu10,Liu10a} has a similar performance. Our proposed eMMSEIA scheme converges to a poorer local optimum value compared to these two schemes. The soft interference nulling (SIN) \cite{NgHuang10} and SDP relaxation \cite{Luo10} algorithms, which use the approximation of the non-convex terms in the objective function, perform worse in this example.
 \par In Fig.~\ref{Fig:2-9}, we evaluate the effect of partial cooperation for the DMMSE, eMMSEIA, and PWF algorithms in a cluster of size $M=5$ where each BS is equipped with $n_t=4$ transmit antennas, each user employs $n_r=2$ receive antennas, and 2 users are dropped randomly in each cell. Recall that the cooperation factor $\kappa$ represents the number of BSs cooperating in transmission to each user. It can be seen that as $\kappa$ increases the performance improves with diminishing returns as $\kappa$ grows large. Moreover, the relative performance of the algorithms confirms the considerations above.
\begin{figure}[t!]
\centering
\includegraphics[width=3.8in]{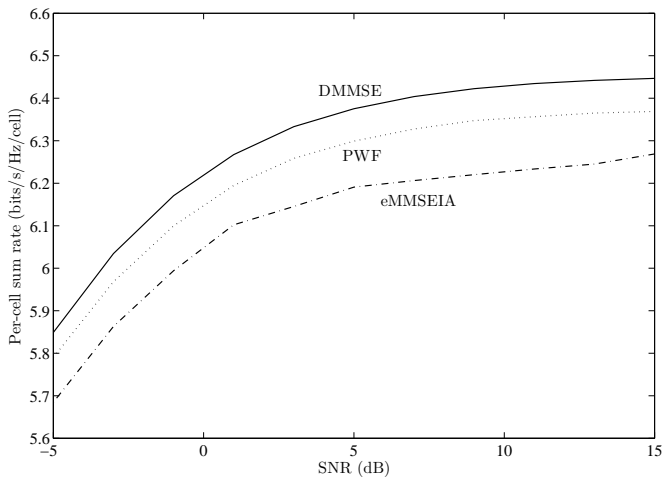}  \caption{Per-cell sum rate of the schemes that can support $d_k < \min(m_{t,k},m_{r,k})$ for $d_k=1$, $n_t=4$, $n_r=2$, $M=3$ and $\kappa=2$.}%
\label{Fig:11}%
\end{figure}
\par In Fig.~\ref{Fig:11}, we compare again the performance of the schemes considered in Fig.~\ref{Fig:2-9} but with a stricter requirement on the number of streams, namely $d_k=1$. It can be seen that the proposed DMMSE tends to perform better than PWF, which was not designed to handle rank constraints. We have adopted the PWF algorithm to support $d_k < \min(m_{t,k},m_{r,k})$ by using a thin SVD of $\mathbf{\hat{\Omega}}_{k}^{-\frac{1}{2}}\mathbf{H}_{k,k}^{\mathsf{H}}\mathbf{\Omega}_{k}^{-\frac{1}{2}}$ when computing (\ref{eqn:pp}).

\begin{figure}[t!]
\centering
\includegraphics[width=3.8in]{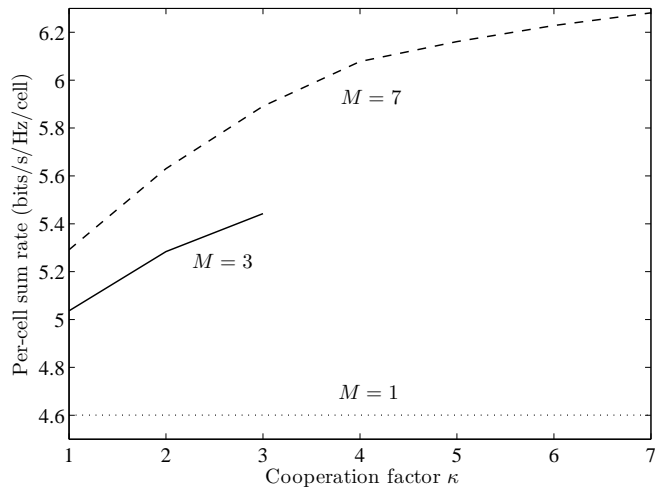}
\caption{Per-cell sum-rate of the proposed DMMSE scheme for cluster sizes $M=1,3,7$ versus the cooperation
factor, $\kappa$, with $n_t=n_r=2$, SNR=20 dB, and single-user per cell.}%
\label{Fig:2-7}%
\end{figure}

\par In Fig.~\ref{Fig:2-7}, we vary the size of the cluster $M$, showing also the advantages of coordinating transmission over larger clusters, even when the number of cooperating BSs $\kappa$ is fixed. Recall that $M$ represents the set of BSs whose transmission is coordinated, but only $\kappa$ BSs cooperate for transmission to a given user.
As an example, for a cluster size of $M=7$ a cooperation factor of $\kappa=4$
performs almost as well as the full cooperation scenario with $\kappa=7$.  Moreover, the performance gains with respect to the non-cooperative case $\kappa=1$ are evident. We also show the performance with a cluster containing a single cell, i.e., $M=1$. This highlights the performance gains attained even in the absence of message sharing among the BSs (i.e., $\kappa=1$) due to the coordination of the BSs within the cluster.

\begin{figure}[t!]
\centering
\includegraphics[width=3.8in]{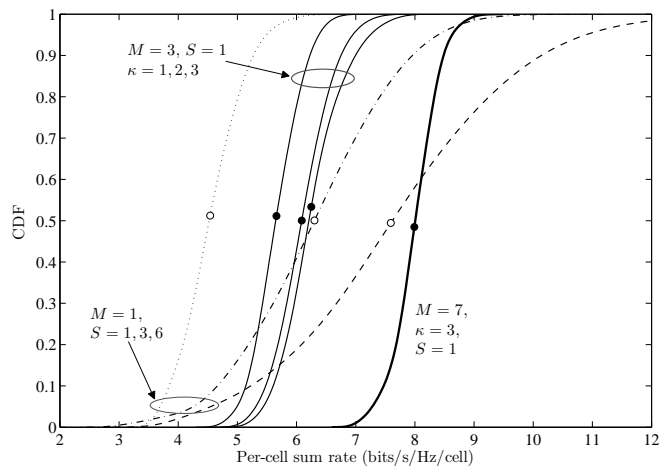}  \caption{CDF of the per-cell sum rates
achieved by DMMSE for $S=1, 3,6$ sectors per cell, $M=1,3,7$ coordinated clusters, and $\kappa=1,2,3$ cooperation factors with $\gamma_0=20 \text{ dB}$, $n_t=6$, and $n_r=2$. The circles represent the mean values of the per-cell sum-rates.}%
\label{Fig:2-10}%
\end{figure}
\par Finally, the effect of sectorization is studied in Fig.~\ref{Fig:2-10} where $n_{t}=6$
transmit antennas at each BS are divided equally into $S=1,3, 6$ sectors. Each cell
contains 6 users, each equipped with $n_r=2$ receive antennas. The users are randomly located at the distance of $\frac{2}{3}d_0$ from its BS. For a given channel realization the DMMSE algorithm is used to obtain the per-cell sum rate. The cumulative distribution functions (CDFs) of per-cell sum rates are computed using large number of channel realizations. The gains of sectorization and cooperation are compared. For example, the system with coordination of 7 cells and $\kappa=3$ cooperation factor and without sectorization performs better than the sectorized system with $S=6$ and without any coordination between the BSs.

\section{Conclusions}

In this paper, we have studied a MIMO interference channel with partial
cooperation at the BSs and per-BS power constraints. We have shown that the
channel at hand is equivalent to a MIMO interference channel under generalized
linear constraints (MIMO-IFC-GC). Focusing on linear transmission strategies,
we have reviewed some of the available techniques for the maximization of the
sum-rate and extended them to the MIMO-IFC-GC when necessary. Moreover, we
have proposed two novel strategies for minimization of the weighted mean
square error on the data estimates. Specifically, we have proposed an extension of the recently introduced MMSE interference alignment strategy and a novel strategy termed diagonalized MSE-matrix (DMMSE). Our proposed strategies support transmission of any arbitrary number of data streams per user. Extensive numerical results show that
the DMMSE outperforms most previously proposed techniques and performs just as well as the best known strategy. Moreover, our results bring insight into the advantages of partial cooperation and sectorization and the impact of the size of the cooperating cluster of BSs and sectorization.

\par We conclude with a brief discussion on the complexity of the algorithms. Due to the difficulty of complete complexity analysis, especially in terms of speed of convergence, we present a discussion based on our simulation experiments. The PWF algorithm converges in almost the same number of iterations as the DMMSE algorithm. The complexity per iteration of PWF and DMMSE is also almost the same as $K\left(\mathcal{O}(\kappa n_tn_r^2)+\mathcal{O}(n_r^3)\right)$ (required for the thin SVD operation). However, the PWF algorithm contains additional operations (matrix inversion and SVD) to obtain the precoding matrices from the calculated transmit covariance matrices.\footnote{This can be performed together with finding the MMSE receive matrices.} Also, the PWF algorithm includes a water-filling algorithm within its inner loop, which is not required in the DMMSE algorithm.
The eMMSEIA algorithm has lower complexity per iteration (i.e. $K\mathcal{O}(n_r^3)$) than the PWF and DMMSE algorithms, since its complexity is due to a matrix inversion per iteration per user. However, eMMSEIA converges in a larger number of iterations than DMMSE and PWF.
The complexity per iteration for the SDP relaxation is higher than for the SIN algorithm (this is because of the extra auxiliary positive semi-definite matrix variable, $\mathbf Y$, introduced in the SDP relaxation algorithm). The SIN algorithm also converges in a smaller number of iterations than the SDP relaxation algorithm.

\appendix[Proof of Lemma 5]
The inequality (\ref{eqn:LowerBound}) follows from weak Lagrangian duality. We now prove the
second part of the statement. Recognizing now that
$\tr\left\{\mathbf{W}\mathbf{E}\right\}$ with (\ref{eqn:MMSE_3})
is a Schur-concave function of the diagonal elements of (\ref{eqn:MMSE_3}%
)\footnote{A Schur-concave function $f(\mathbf{x})$ of vector $\mathbf{x=(}%
x_{1},...,x_{d})$ is such that $f(\mathbf{x})\leq f(\mathbf{x}^{\prime})$ if
$\mathbf{x}$ majorizes $\mathbf{x}^{\prime},$ that is, if $\sum_{i=1}%
^{j}x_{[i]}\geq\sum_{i=1}^{j}x_{[i]}^{\prime}$ for all $j=1,...,d,$ where
$x_{[i]}$ (and $x_{[i]}^{\prime})$ represents the vector sorted in decreasing
order, i.e., $x_{[1]}\geq...\geq x_{[d]}$ (and $x_{[1]}^{\prime}\geq...\geq
x_{[d]}^{\prime}).$}, it can be argued that the minimum is attained when
$\mathbf{E}$ is diagonalized as we did for Lemma 4. Defining $\mathbf R=\mathbf H^\mathsf H \mathbf \Omega^{-1}\mathbf H$, we can conclude that $\mathbf{B}^{\mathsf{H}}\mathbf{R}\mathbf{B}$ must be also diagonal in this search domain.
Now assume that an optimal solution of the single-user WSMMSE problem is denoted as $\tilde
{\mathbf{B}}$. Without loss of generality we can assume that this solution
diagonalizes the MSE matrices. 
The necessity of the KKT conditions can be proved as in \cite{Liu10a} and in special cases such as the MIMO interference channel with partial message sharing of Sec. II, it also follows from linear independence constraint qualification conditions \cite{Bertsekas03}.

Hence, there exists a Lagrange multiplier vector $\tilde{\boldsymbol{\lambda}}$ which together with
$\tilde{\mathbf{B}}$ satisfies the KKT conditions of the WSMMSE problem (\ref{eqn:optimization22}) \cite{Christensen08}\cite{Bertsekas03}. As it is stated in the Lemma, we consider the case that $\tilde{ \lambda}_m $ are also strictly positive (i.e. $\tilde{\lambda}_m >0$ for all $m$). Simplifying the KKT condition (\ref{eqn:KKTcond1}), we have\footnote{We use differentiation rule $\nabla_{\mathbf X} \tr\left\{\mathbf A\mathbf X^\mathsf H\mathbf B\right\}=\mathbf B\mathbf A$ and $\nabla_{\mathbf X}\tr\left\{\mathbf Y^{-1}\right\}=-\mathbf Y^{-1}\left(\nabla_{\mathbf X}\mathbf Y\right) \mathbf Y^{-1}$. For the complex gradient operator each matrix and its conjugate transpose are treated as independent variables \cite{Hjorugnes07}.}
\begin{equation}
\nabla_{\mathbf B}\mathcal{L}=-\mathbf{R}\tilde{\mathbf{B}}\tilde{\mathbf{E}}%
\mathbf{W}\tilde{\mathbf{E}}+\sum\limits_{m=1}^{M}\tilde{\lambda
}_{m}\mathbf{\Phi}_{m}\tilde{\mathbf{B}}  =\mathbf{0}%
\label{eqn:KKTcond1_simplified1}
\end{equation}
Left-multiplying (\ref{eqn:KKTcond1_simplified1}) by $\tilde{\mathbf{B}}^{\mathsf{H}}$
gives us
\begin{equation}
\tilde{\mathbf{B}}^{\mathsf{H}}\mathbf{R}\tilde{\mathbf{B}}%
\tilde{\mathbf{E}}\mathbf{W}\tilde{\mathbf{E}}=\tilde{\mathbf{B}%
}^{\mathsf{H}}\left(  \sum_{m}\tilde{\lambda}_{m}\mathbf{\Phi}%
_{m}\right)  \tilde{\mathbf{B}}.\label{eqn:KKTcond1_simplified}%
\end{equation}
Since $\tilde{\mathbf{B}}^{\mathsf{H}}\mathbf{R}\tilde{\mathbf{B}}%
$ and correspondingly $\tilde{\mathbf{E}}$ are diagonal matrices,
$\tilde{\mathbf{B}}^{\mathsf{H}}\left(  \sum_{m}\tilde{\lambda}%
_{m}\mathbf{\Phi}_{m}\right)  \tilde{\mathbf{B}}$ must also be diagonal.
For simplicity, we introduce $\mathbf{\Phi}(\tilde{\boldsymbol{\lambda
}})=\sum_{m=1}^{M}\tilde{\lambda}_{m}\mathbf{\Phi}_{m}$. Since $\tilde{\lambda}_m>0$ for every $m$, therefore $\mathbf \Phi(\tilde{\boldsymbol \lambda})$ is a non-singular matrix. This can be easily verified due to the structure of $\mathbf \Phi_m$.
Hence, we can write $\tilde{\mathbf{B}}^{\mathsf{H}}\mathbf{\Phi}%
(\tilde{\boldsymbol{\lambda}})\tilde{\mathbf{B}}=\tilde{\mathbf{\Delta}%
}$ where $\tilde{\mathbf{\Delta}}\in \mathbb{C}^{d\times d}$ is a
diagonal matrix. Therefore, we can write
\begin{equation}
\mathbf \Phi(\tilde{\boldsymbol \lambda})^{1/2}\tilde{\mathbf B}=\tilde{\mathbf U} \tilde{\mathbf \Sigma}
\end{equation}
where $\tilde{\mathbf U} \in \mathbb C^{m_{t} \times d}$ consists of orthonormal columns (i.e. $\tilde{\mathbf U}^\mathsf H \tilde{\mathbf U}$) and $\tilde{\mathbf \Sigma} \in \mathbb C^{d \times d}$ is a diagonal matrix with the diagonal terms of $\sqrt{\tilde{p}_{i}}$. Hence, we can write
\begin{equation}\label{eqn:optimalB_k_su_mc}
\tilde{\mathbf B}=\mathbf \Phi(\tilde{\boldsymbol \lambda})^{-1/2}\tilde{\mathbf U} \tilde{\mathbf \Sigma}.
\end{equation}
Replacing the structure of $\tilde{\mathbf B}$ given in (\ref{eqn:optimalB_k_su_mc}), we can write
\begin{equation}
\tilde{\mathbf{B}}^{\mathsf{H}}\mathbf{R}\tilde{\mathbf{B}}=\tilde{\mathbf \Sigma}^\mathsf H\tilde{\mathbf U}^\mathsf H\mathbf \Phi(\tilde{\boldsymbol \lambda})^{-\frac{1}{2}}\mathbf R \mathbf \Phi(\tilde{\boldsymbol \lambda})^{-\frac{1}{2}}\tilde{\mathbf U} \tilde{\mathbf \Sigma}=\mathbf D
\end{equation}
Thus, we can conclude from the equation above that $\tilde{\mathbf U}$ must contain the eigenvectors of $\mathbf \Phi(\tilde{\boldsymbol \lambda})^{-\frac{1}{2}}\mathbf R \mathbf \Phi(\tilde{\boldsymbol \lambda})^{-\frac{1}{2}}$.
\par Now, plugging (\ref{eqn:optimalB_k_su_mc}) into (\ref{eqn:KKTcond1}) and left-multiply it with $\mathbf \Phi^{-\frac{1}{2}}$, we get
\begin{equation}
\tilde{\mathbf \Gamma} \tilde{\mathbf{\Sigma}} \left( \mathbf{I}%
+ \tilde{\mathbf \Gamma}\tilde{\mathbf{\Sigma}}^{2}\right)
^{-1}\mathbf{W} \left( \mathbf{I}+
\tilde{\mathbf \Gamma}\tilde{\mathbf{\Sigma}}^{2}\right) ^{-1}=\tilde{\mathbf{\Sigma}}%
\label{eqn:Gamma}
\end{equation}
where $\tilde{\mathbf \Gamma}(\tilde{\boldsymbol \lambda})=\diag[\gamma_{1}(\tilde{\boldsymbol \lambda}) \cdots \gamma_{d}(\tilde{\boldsymbol \lambda})]$ is a diagonal matrix with the diagonal terms of the $d$ largest eigenvalues of $\mathbf \Phi(\tilde{\boldsymbol \lambda})^{-\frac{1}{2}}\mathbf R \mathbf \Phi(\tilde{\boldsymbol \lambda})^{-\frac{1}{2}}$.
Since all the matrices are diagonal, (\ref{eqn:Gamma}) reduces to the scalar equations:
\begin{equation}
\frac{w_{i}\gamma_{i}(\tilde{\boldsymbol \lambda})%
}{(1+\tilde{p}_{i}\gamma_{i}(\tilde{\boldsymbol \lambda}))^{2}}%
=1%
\end{equation}
Solving these equations gives us the optimal $\tilde{p}_{i}$ given by
\begin{equation}
\label{eqn:opt_pk_sumc}\tilde{p}_{i}=\left[ \sqrt{\frac{w_{i}}%
{\gamma_{i}(\tilde{\boldsymbol{\lambda}})}}-\frac{1}{\gamma_{i}(\tilde{\boldsymbol{\lambda}})}\right] ^{+}, %
\end{equation}
 Thus, for the given Lagrange multiplier $\tilde{\boldsymbol \lambda}$ which together with $\tilde{\mathbf B}$, satisfying the KKT conditions of (\ref{eqn:optimization22}), $\tilde{\mathbf B}$ must satisfy (\ref{eqn:optimalB_k_su_mc}) and (\ref{eqn:opt_pk_sumc}). If all power constraints are satisfied with equality by this solution, then (\ref{eqn:optimalB_k_su_mc}) and (\ref{eqn:opt_pk_sumc}) also solves the single constraint problem
\begin{equation}%
\begin{array}
[c]{rl}%
\underset{\mathbf{B}}%
{\text{minimize}} & \tr\left\{\mathbf W\left(\mathbf I+\mathbf B^\mathsf H \mathbf H^\mathsf H \mathbf \Omega^{-1}\mathbf H \mathbf B\right)^{-1}\right\}\\
\text{subject to} & \tr\left\{  \mathbf{\Phi}(\tilde{\boldsymbol \lambda})%
\mathbf{B}\mathbf{B}^{\mathsf{H}}\right\}  \leq \sum\limits_{m=1}^M\tilde{\lambda}_mP_{m},
\end{array}
. \label{eqn:optimization33}%
\end{equation}
The solution of this problem is given in Lemma 3 as
\begin{equation}
\label{eqn:Bk_P2_2}\mathbf{B}(\tilde{\boldsymbol{\lambda}})=\mathbf \Phi(\tilde{\boldsymbol \lambda})^{-\frac{1}{2}}\mathbf U \mathbf \Sigma%
\end{equation}
where $\mathbf U$ consists of $d$ eigenvectors of $\mathbf \Phi(\tilde{\boldsymbol \lambda})^{-\frac{1}{2}}\mathbf R\mathbf \Phi(\tilde{\boldsymbol \lambda})^{-\frac{1}{2}}$ corresponding to its largest eigenvalues and $\mathbf{\Sigma}$ is a diagonal matrix with the diagonal elements of $\sqrt{p_{i}}$, which is
given by
\begin{equation}
\label{eqn:opt_pk_susc_2}p_{k,i}=\left[ \sqrt{\frac{w_{i}}{\mu\gamma
_{i}(\tilde{\boldsymbol{\lambda}})}}-\frac{1}{\gamma
_{i}(\tilde{\boldsymbol{\lambda}})}\right] ^{+},%
\end{equation}
for a waterfilling value of $\mu\geq0$ which satisfies the power constraint
\begin{equation}
\label{eqn:power_Cons_P2}\tr\left\{  \mathbf \Phi(\tilde{\boldsymbol \lambda}) \mathbf{B}(\tilde{\boldsymbol{\lambda}%
})\mathbf{B}(\tilde{\boldsymbol{\lambda}})^{\mathsf{H}}\right\}  \leq
\sum_{m} \tilde{\lambda}_{m}P_{m}.
\end{equation}
On the other hand, summing up the KKT conditions $\tilde{\lambda}_m\left(P_m-\tr\left\{\mathbf \Phi_m\mathbf B\mathbf B^\mathsf H\right\}\right)=0$
for all $m$, we obtain that
\begin{equation}
\tr\left\{  \left( \sum_{m}\tilde{\lambda}_{m}\mathbf{\Phi}_{m}\right)
\tilde{\mathbf{B}}\tilde{\mathbf{B}}^{\mathsf{H}}\right\}  = \sum_{m}
\tilde{\lambda}_{m}P_{m}%
\end{equation}
If we set $\mu=1$ and comparing (\ref{eqn:opt_pk_sumc}) and
(\ref{eqn:opt_pk_susc_2}), we can conclude that $\tilde{p}_{i}=p_{i},
\forall i$ which together with comparison of (\ref{eqn:Bk_P2_2}) and
(\ref{eqn:optimalB_k_su_mc}) we can conclude that $\mathbf{B}(\tilde
{\boldsymbol{\lambda}})=\tilde{\mathbf{B}}$ and the $\mu=1$ is the optimal Lagrange multiplier of the single-constraint WSMMSE problem (\ref{eqn:optimization33}). Following Lemma 4, this precoding matrix is also a result of minimization of the Lagrangian function (\ref{eqn:Lagrangian_su}) when $\mu=1$ and $\mathbf \Phi_1=\mathbf \Phi(\tilde{\boldsymbol \lambda})$, which means
\begin{equation}\label{eqn:LowerBound2}
 p^\star=\inf_{\mathbf B}\mathcal L(\mathbf B;\tilde{\boldsymbol \lambda}).
 \end{equation}
  On the other hand, we have
\begin{equation}\label{eqn:LowerBound3}
\max_{\boldsymbol \lambda \geq 0}\inf_{\mathbf B}\mathcal L(\mathbf B;\boldsymbol \lambda )\geq \inf_{\mathbf B}\mathcal L(\mathbf B;\tilde{\boldsymbol \lambda})
\end{equation}
which in concert with (\ref{eqn:LowerBound}) and (\ref{eqn:LowerBound2}) results in
\begin{equation}
p^\star=\inf_{\mathbf B}\mathcal L(\mathbf B;\tilde{\boldsymbol \lambda})=\max_{\boldsymbol \lambda \geq 0}\inf_{\mathbf B}\mathcal L(\mathbf B;\boldsymbol \lambda ),
\end{equation}
thus concluding the proof.

\bibliographystyle{IEEEtran}
\bibliography{IEEEabrv,MyBib}
\end{document}